\documentclass[fleqn,usenatbib]{mnras}

\usepackage{mathptmx}

\usepackage[T1]{fontenc}
\usepackage{ae,aecompl}
\usepackage{longtable}

\usepackage{color}
\usepackage{units}
\usepackage{epstopdf}
\usepackage{hyperref}
\usepackage{multirow}
\usepackage{float}
\usepackage{tikz}

\usetikzlibrary{shapes,arrows}

\usepackage{listings}
\usepackage{color}

\usepackage[titletoc,title]{appendix}

\usepackage{graphicx}	
\usepackage{amsmath}	
\usepackage{amssymb}	
\usepackage{booktabs,siunitx}

\title[Predicting the electromagnetic counterparts using low-latency, gravitational-wave data products]{Predicting electromagnetic counterparts using low-latency, gravitational-wave data products}

\author[Stachie et al.]{
Cosmin Stachie$^{1}$\thanks{E-mail: scosmin@oca.eu}, 
Michael W. Coughlin$^{2}$,
Tim Dietrich$^{3,4}$,
Sarah Antier$^{5}$,
Mattia Bulla$^{6}$, \newauthor
Nelson Christensen$^{1}$, 
Reed Essick$^{7,8}$,
Philippe Landry$^{9}$, 
Benoit Mours$^{10}$,\newauthor
Federico Schianchi$^{3}$,
Andrew Toivonen$^{2}$
\\
$^{1}$Artemis, Universit\'e C\^ote d'Azur, Observatoire C\^ote d'Azur, CNRS, CS 34229, F-06304 Nice Cedex 4, France \\
$^{2}$School of Physics and Astronomy, University of Minnesota, Minneapolis, Minnesota 55455, USA\\
$^{3}$Institut f\"{u}r Physik und Astronomie, Universit\"{a}t Potsdam, Haus 28, Karl-Liebknecht-Str. 24/25, 14476, Potsdam, Germany \\
{${}^4$Max Planck Institute for Gravitational Physics (Albert Einstein Institute), Am M\"uhlenberg 1, Potsdam 14476, Germany}\\
$^{5}$Universit\'{e} de Paris, CNRS, Astroparticule et Cosmologie, F-75013 Paris, France \\
$^{6}$The Oskar Klein Centre, Department of Astronomy, Stockholm University, AlbaNova, SE-106 91 Stockholm, Sweden\\
$^{7}$Kavli Institute for Cosmological Physics, The University of Chicago,5640 South Ellis Avenue, Chicago, Illinois, 60637, USA \\
$^8$Perimeter Institute for Theoretical Physics, 31 Caroline Street North, Waterloo, Ontario, Canada, N2L 2Y5 \\
$^{9}$Gravitational-Wave Physics \& Astronomy Center, California State University, Fullerton, 800 N State College Blvd, Fullerton, CA 92831\\
$^{10}$Institut Pluridisciplinaire Hubert CURIEN, 23 rue du loess - BP28 67037 Strasbourg cedex 2, France
}

\date{Accepted XXX. Received YYY; in original form ZZZ}

\pubyear{2021}

\begin{document}
\label{firstpage}
\pagerange{\pageref{firstpage}--\pageref{lastpage}}
\maketitle

\begin{abstract}

Searches for gravitational-wave counterparts have been going in earnest since GW170817 and the discovery of AT2017gfo.
Since then, the lack of detection of other optical counterparts connected to binary neutron star or black hole - neutron star candidates has highlighted the need for a better discrimination criterion to support this effort.
At the moment, the low-latency gravitational-wave alerts contain preliminary information about the binary properties and, hence, on whether a detected binary might have an electromagnetic counterpart. The current alert method is a classifier that estimates the probability that there is a debris disc outside the black hole created during the merger as well as the probability of a signal being a binary neutron star, a black hole - neutron star, a binary black hole or of terrestrial origin.
In this work, we expand upon this approach to predict both the ejecta properties and provide contours of potential lightcurves for these events in order to improve follow-up observation strategy. The various sources of uncertainty are discussed, and we conclude that our ignorance about the ejecta composition and the insufficient  constraint of the binary parameters, by the low-latency pipelines, represent the main limitations.
To validate the method, we test our approach on real events from the second and third Advanced LIGO-Virgo observing runs.
\end{abstract}




\section{Introduction}

The search for, detection, and the characterization of the \emph{kilonova} AT2017gfo \citep{CoFo2017,SmCh2017,AbEA2017f}, associated with the binary neutron star (BNS) merger GW170817 \citep{AbEA2017b} and the short gamma-ray burst GRB170817A~\citep{Goldstein:2017mmi, Savchenko:2017ffs, Abbott_2017},  has spurred on the search for more of these objects. These kilonovae are expected to be produced in many of the mergers of compact objects involving at least one neutron star (with another neutron star or black hole as companion).
Powered by the neutron-rich outflows undergoing the radioactive decay of r-process elements \citep{LaSc1974,LiPa1998,MeMa2010,KaMe2017}, these ultra-violet/optical/infrared transients produce emissions approximately isotropically\footnote{We emphasize that despite the approximately isotropic nature of the kilonvoae, an angular dependence exists as pointed out in, e.g., \citep{Perego:2017wtu,Kawaguchi:2019nju,Heinzel:2020qlt}.} and therefore are visible from nearly all directions.
The properties of the kilonova, including the lightcurves and spectra, depend on the parameters of the original binary, including the masses (typically characterized by the chirp mass and mass ratio), spin angular momentum, and the equation of state describing the neutron stars' interior \citep{BaBa2013,PiNa2013,AbEA2017b,BaJu2017,DiUj2017,RaPe2018}.
The association between lightcurves and binary parameters has been used to place constraints on the character of the progenitor systems and quantity of matter expelled, e.g., \citep{KaMe2017,CoDi2017,SmCh2017,Perego:2017wtu,Hinderer:2018pei,Kawaguchi:2019nju, Bulla:2019muo,Coughlin:2019kqf, 2021arXiv210202229N,Raaijmakers:2021slr}.

Searches for these counterparts are difficult for a variety of reasons, the most important one is the large sky localizations spanning $\approx 100-10,000\,\textrm{deg}^2$ \citep{Rover2007a, Fair2009,Fair2011,Grover:2013,WeCh2010,SiAy2014,SiPr2014,BeMa2015,EsVi2015,CoLi2015,KlVe2016}.
Due to the size of the localizations, wide-field survey telescopes such as the Panoramic Survey Telescope and Rapid Response System (Pan-STARRS; \citep{MoKa2012}), Asteroid Terrestrial-impact Last Alert System (ATLAS; \citep{ToDe2018}), the Zwicky Transient Facility (ZTF; \citep{Bellm2018,Graham2018,DeSm2018,MaLa2018}), telescope networks such as the Gravitational-Wave Optical Transient Observer (GOTO-4) \citep{GoCu2020}), Global Rapid Advanced Network Devoted to the Multi-messenger Addicts (GRANDMA \citep{GRANDMA2020, Antier:2019pzz})  and future facilities such as BlackGEM \citep{BlGr2015} and the Vera C.\ Rubin Observatory's Legacy Survey of Space and Time (LSST; \citep{Ivezic2014}), can most efficiently cover the extended regions.

Given limited telescope time, prioritization of gravitational-wave (GW) event candidates for follow-up is essential. This can include considerations such as the false alarm rate of the event, the time of the merger (and therefore its relation to observability~\citep{Chen2017}) and properties of the merger itself. In particular, one quantity of interest is the apparent magnitude of the lightcurve in bands of a particular telescope during its observability window. This would limit observations to those objects that could be feasibly detected given the available telescope time and would help prioritizing between exposure time and sky coverage. An observation strategy, based on the idea of using low-latency GW products to predict electromagnetic (EM) properties, was first introduced in~\citep{SoCo2017}. \\

A number of previous studies tried to address the question of which compact binary merger should be the target of EM observations, e.g., \cite{Pannarale:2014rea} were one of the first who used the remnant matter outside the final black hole as a proxy for the likelihood of potential EM counterparts. In addition, based on general-relativistic numerical simulations, empirical fitting formulas have also been derived for the ejected material and for the disc mass for BNS systems \citep{DiUj2017,Radice:2018pdn,CoDi2018,Dietrich:2020lps,Nedora:2020qtd} and black hole-neutron star (BHNS) systems \citep{Foucart:2012nc,Kawaguchi:2016ana,Foucart:2018rjc,KrFo2020}.

Rapid analysis of the GW data in the era of the advanced detectors is done by online low-latency pipelines. Traditionally there are two types of pipelines: one category targeting modeled signals and the other category tracking unmodeled events, both signals being General Relativity predictions. Thereby the pipelines of the first type search for well-modeled predicted signals~\citep{Canton:2014ena, 2020arXiv201005082C,Aubin:2020,Hooper:2011rb}, whereas the other type of pipelines searches for an excess of power in the data~\citep{Klimenko:2008fu, PhysRevD.95.104046, PhysRevD.103.044006, Sutton_2010}. For the present study, we will use the templates which came out at the end of an analysis realised by the multi-band template analysis (MBTA) pipeline~\citep{Aubin:2020}, which searches for modeled binary mergers.

The realtime public data products~\citep{article} to aid the EM/neutrino follow-up of binary merger candidates include 3D sky localization \citep{SiPr2016,Si2016}, the probability that the candidate is an astrophysical event \citep{Kapadia:2019uut}, the probability of having at least one neutron star --chraracterized by the probability of having one companion with mass below $3M_\odot$ -- and the probability of having remnant matter from the merger \citep{Chatterjee:2019avs} -- based on the disc mass prediction of~\cite{Foucart:2018rjc}.  Overall, while extremely useful, it requires that all compact objects with masses below $3M_\odot$ to be neutron stars, and  there are some shortcomings to this analysis, e.g., not all BNS mergers will have a detectable EM counterpart, \citep[e.g.][]{Coughlin:2019zqi,Coughlin:2020fwx,Bauswein:2020xlt}. A source classifier based on the template chirp mass was equally discussed in~\cite{2020arXiv200807494D}. Likewise, information from presumable compact binary coalescence EM precursors~\citep{2021MNRAS.501.3184S, Schnittman:2017nhg} might be envisaged in the future.

One issue to overcome, in addition to the statistical uncertainties, is the systematic errors in the low-latency template based analysis. These searches use discrete template banks of waveforms to perform matched filtering on the data. For the online searches, which are what we will be concerned here, the templates are characterized by masses, $m_1$ and $m_2$, and the dimensionless aligned/anti-aligned spins of the binary elements along the orbital angular momentum of the binary, $s_1$ and $s_2$. These pipelines report the best matching templates based on a detection statistic, giving a point estimate of these four quantities. The downside to this is clear: while quantities like the \emph{chirp mass} $m_\mathrm{chirp}$ of systems are well measured, mass ratio and spin tend to be poorly constrained by this point estimate~\citep{Biscoveanu_2019}.

Additional, important supra-nuclear matter equation of state dependent information not provided by the low-latency pipelines are estimates of the maximum mass, compactness and/or tidal deformability of neutron stars. The maximum mass informs the classification of events as BNS, neutron star - black hole (NSBH), or binary black hole (BBH)~\citep{ Essick:2020ghc}. 

The presence or absence of an EM counterpart to a compact binary coalescence is determined by the amount of unbound baryonic material. The amount of ejecta, or even whether there is measurable ejecta, is directly linked either to the compactness of the neutron star(s) or to their tidal deformability $\Lambda$ in the combination
\begin{equation}
     \tilde\Lambda = \frac{16}{13}\frac{(m_1 + 12m_2)m_1^4\Lambda_1 + (m_2 + 12m_1)m_2^4\Lambda_2}{(m_1 + m_2)^5}.
\end{equation}
In general, the larger the tidal deformability, the less compact the stars and the higher the probability of gravitationally unbound material producing bright kilonovae. 

In order to create a prior for the compactness $C$  and maximum neutron-star mass, a choice of the neutron star equation of state is necessary. The equations of state employed in this work are a zero temperature relation between pressure and the rest-mass energy density governing a fluid of baryons at supra-nuclear densities. Given an equation of state, there is a one-to-one correspondence between mass and tidal deformability, if the neutron star is completely made up of hadrons. Indeed, in the case of hybrid stars, hypothetical objects where deconfined quarks might exist~\citep{Alford:2013aca, Han:2019bub, PhysRevD.58.024008}, the situation is different. Hybrid equations of state can support twin stars, neutron stars with the same mass but different central densities: the lower-density star’s core is hadronic, while the higher-density star’s is quark-like~\citep{Essick:2019ldf, Chatziioannou:2019yko,Pang:2020ilf}. For this study we consider only the case of hadron stars. Moreover, the supra-nuclear matter equation of state is important for the determination of a maximum neutron star mass. Effectively a soft (stiff) equation of state means more (less) compact neutron stars corresponding to lower (higher) maximum mass. Unfortunately the supra-nuclear matter equation of state is not known exactly despite progress by different methods: simultaneous measurement of neutron star mass and radius, e.g.,~\citep{Miller:2019cac, Lattimer_2001,Miller:2019cac,Raaijmakers:2019dks,Riley:2019yda,Bogdanov:2019ixe}; 
combination of gravitational tidal effect and EM data, e.g.,~\citep{Radice:2018ozg,Dietrich:2020lps,PhysRevD.101.123007,Breschi:2021tbm}; 
or by a combination of nuclear physics and multi-messenger astronomy observations, e.g., ~\citep{Capano:2019eae,Dietrich:2020lps, PhysRevC.102.055803}. 

As stated above, the mass ejecta is a key ingredient in the derivation of kilonova lightcurves. However, numerical simulations relying on General Relativity are required to estimate this quantity. Despite the existence of such calculations~\citep[e.g.,][]{Goriely:2011vg,Rosswog:2013kqa, Grossman:2013lqa,Tanaka:2013ana,Dietrich:2018phi,Radice:2018pdn,Bovard:2017mvn,Shibata:2017xdx,Foucart:2019bxj}, they are computationally expensive, and cannot be performed directly in the minutes following a GW alert. For this reason, groups have proposed fits for the ejecta mass based on numerical-relativity simulations for both BNS mergers, e.g.,~\citet{DiUj2017,Radice:2018pdn,CoDi2018,Dietrich:2020lps,Nedora:2020qtd}, and NSBH mergers, e.g.,~\citet{Foucart:2012nc,Kawaguchi:2016ana,Foucart:2018rjc,KrFo2020}. The present paper aims to put together such existing tools as well as  parametrized kilonova lightcurve models~\citep{KaMe2017, Bulla:2019muo} in order to predict EM counterparts based on only low-latency GW pipeline signal-to-noise distributions over the template bank.     

The remainder of the paper is structured as follows: In Section~\ref{sec:MBTA_uncertainties}, we discuss the GW low-latency analysis and the current parameters released to aid observers. Section~\ref{sec:ejected_matter} presents how we convert component binary parameters to mass ejecta and we discuss the two models that we employ in the computation of kilonova lightcurves in Section~\ref{sec:surrogates}. We validate our method on GW events from recent LIGO-Virgo Observing Runs in Section~\ref{sec:O2/O3 events}. We summarize the performance of this tool and suggest improvements for future work in Section~\ref{sec:conclusion}.

\section{Addressing the point estimate uncertainties}
\label{sec:MBTA_uncertainties}
MBTA ~\citep{Aubin:2020} is a modeled search pipeline based on matched filtering, which compares the inspiral waveforms from a  ``bank'' of the templates to the data. 
Templates are distributed across the parameter space such that any point has a good match with at least one of the templates of the bank, the minimal match value being typically 97\% (for GW170817 we used here 99\%). 
The template bank is therefore a rather uniform sampling of the parameter space.
This template bank is applied separately to each detector; coincident triggers are those that share the same template parameters and have time delays consistent with astrophysical sources. 

MBTA splits this analysis in two or more frequency bands, i.e., instead of comparing all the frequency components of the data to those of the template, the frequency band of the detector data and templates are split into multiple bands.\footnote{Other pipelines also perform multi-band analyses, an example being~\citet{Sachdev:2019vvd}.} The matched filter is computed within each band, and the signal-to-noise ratio corresponding to the different bands are combined to assign an overall statistical significance to the template. 
This procedure reduces the computational cost such that the pipeline is able to analyze the LIGO-Virgo data with a sub-minute latency using modest computing resources (about 150 cores). 
It is worth mentioning the analysis pursued in this work should apply equally well to all low-latency pipelines.

\subsection{Template uncertainties}
As mentioned previously, during observing runs O2 and O3, the low-latency alerts released by the LIGO Scientific Collaboration and Virgo Collaboration (LVC) consisted of the binary parameters of the template with the highest statistical significance. In Table \ref{tab:preferred_events}, these parameters are displayed for GW170817, GW190425, and GW190814.  
The reason we focus on these events is that they are the unambiguously confirmed binary systems which have a non-negligible probability to possess at least one neutron star. 
However, for the present work, we consider not only the ``best'' template, with corresponding $\rm SNR_{\rm max}$, but all templates with signal-to-noise ratio,  $\rm SNR > SNR_{\rm max} - 3$, 
$\rm SNR\textsubscript{max}$ 
is the signal-to-noise ratio of the ``best'' template. The motivation for this choice is the desire to realize a rapid parameter estimation based on these neighbourhood templates
which are within three standard deviations of the best template and therefore capture 99.7\% of the parameters information.

\begin{table*}
\caption[]{The MBTA preferred templates that maximize the signal-to-noise ratio for GW170817, GW190425, and GW190814.
We include the name of the event, the mass of the more massive compact object, $m_1$, the mass of the lighter compact object, $m_2$, the chirp mass $m_\mathrm{chirp}$, the mass ratio $q$, the projection of the heavier binary component's spin in the direction of the orbital angular momentum $s_1$, the projection of the lighter binary component's spin in the direction of the orbital angular momentum $s_2$.}
\label{tab:preferred_events}
\begin{tabular}{lcccccccc}
Event & $m_1$ & $m_2$ & $m_\mathrm{chirp}$ & $q$ & $s_1$ & $s_2$ & $\chi_\mathrm{eff}$ \\ 
 & $(M_{\odot})$ &$(M_{\odot})$ & $(M_{\odot})$ & & & & \\
\hline
GW170817 & 1.674 & 1.139 & 1.198 & 0.680  & 0.040 & 0.000 & 0.024 \\ 
GW190425 & 2.269 & 1.305 & 1.487 & 0.575 & 0.080 & -0.010 & 0.047 \\ 
GW190814 & 36.881 & 2.093 & 6.522 & 0.057 & 0.340 & 0.960 & 0.373 \\ 
\hline
\end{tabular}
\end{table*}

\subsection{Using multiple templates}
\label{sec:multiple_templates}

For one event, MBTA provides the list of templates that have been triggered, with their SNR. 
For each of them, a weight, $w$, is given to capture the probability that this template is the most likely to describe the event. 
It is based on the SNR of the template $i$ relative to the maximum SNR: dSNR = SNR$_{max} $ – SNR$_i$, that is the number of standard deviations for this template compared to the best template.
The weights are computed by sorting the templates by increasing dSNR, and then getting the difference of the error function with the following template:
$w_i$ = erf($\mathrm{dSNR}_{i+1}/\sqrt{2}$) - erf($\mathrm{dSNR}_i/\sqrt{2}$). 
Before being used, the weights are smoothed by averaging them with their two adjacent templates in dSNR. 
With this procedure, the sum of all weights is one.
We will use the weights as the ``significance'' measure for a given template.

The input data is represented by a list of templates, which is a 5-tuple $(m_1, m_2, s_1, s_2, w)$, where $m_1, m_2$ ($m_1 \geq m_2$) are the masses of the binary components, $s_1, s_2$ are the projections of the spins onto the direction of the orbital angular momentum, and $w$ is the normalized weight. 
In Table~\ref{tab:table_limits} we list the median, lower and upper limits for the binary parameters obtained by this procedure. The corresponding values obtained by the more expensive offline parameter estimation method~\citep{Veitch:2014wba} (hereafter PE) are also presented. One can observe that there is a very good similarity (at most a few percent deviation) between PE and our method for $m_\mathrm{chirp}$. On the other hand the mass ratio and effective spin distributions can be very different (more than 100\% in the case of GW190814). One could imagine different ways to address the problem of these latter distributions. A possibility might be to consider a population prior based on the already detected binary compact merger events as suggested in, e.g.,~\citet{Essick:2020ghc,Fishbach:2020ryj,Fishbach_2020,PhysRevD.81.084029, Abbott:2020gyp}, however this procedure might introduce additional biases if yet unobserved populations of compact binaries exist. 

\begin{table*}
  \caption{The median, the upper limits (90th percentile) and the lower limits (10th percentile) for GW170817, GW190425 and GW190814 parameters for the chirp mass $m_\mathrm{chirp}$, the mass ratio $q$, and the effective spin $\chi_{\rm eff}$. These quantities are obtained from the set of templates generated as explained in Section~\ref{sec:multiple_templates}. The same quantities obtained from offline PE posteriors are also illustrated for comparison. A missing $m_\mathrm{chirp}$ upper/lower limit means that the deviation from the median value is less than $0.001 M_\odot$. The PE samples have been introduced previously~\citep{LIGOScientific:2018mvr, Abbott:2020niy}.}
  \label{tab:table_limits}
\renewcommand{\arraystretch}{1.35}
    \begin{tabular}{l|ccc|ccc}
             & \multicolumn{3}{c|}{MBTA} & \multicolumn{3}{c}{PE} \\
\hline
 Event & $m_\mathrm{chirp}$  & ${q}$ &  $\chi_{\rm eff}$ &  $m_\mathrm{chirp}$  & ${q}$ &  $\chi_{\rm eff}$ \\
  & $(M_{\odot})$  &  &   & $(M_{\odot})$ & &  \\ \hline
GW170817 & $1.198_{-0.001}$ & $0.756_{-0.157}^{+0.068}$ & $0.029_{-0.018}^{+0.017}$ & $1.186_{\phantom{+0.001}}$ & $0.864_{-0.12}^{+0.107}$ & $0.003_{-0.007}^{+0.01}$ \\
GW190425 & $1.487_{-0.002}^{+0.001}$ & $0.784_{-0.229}^{+0.121}$ & $0.026_{-0.056}^{+0.024}$ & $1.437_{-0.016}^{+0.018}$ & $0.657_{-0.21}^{+0.266}$ & $0.058_{-0.041}^{+0.079}$ \\
GW190814 & $6.474_{-0.134}^{+0.125}$ & $0.058_{-0.008}^{+0.178}$ & $0.321_{-0.817}^{+0.094}$ & $6.09_{-0.043}^{+0.046}$ & $0.111_{-0.007}^{+0.006}$ & $-0.003_{-0.045}^{+0.047}$ \\
\hline
\end{tabular}
\label{tab:addlabel}
\end{table*}


In the following, the intrinsic masses and spins estimated here will be used for the computation of the mass and the velocity of ejecta in Section~\ref{sec:ejected_matter}. It is worth mentioning that over the past years several rapid parameter estimation efforts have been realized, e.g., \citep{Pankow:2015cra, Lange:2018pyp,Smith:2019ucc}. 

\section{Dynamical and disc wind ejecta from templates}
\label{sec:ejected_matter}
Two important features of a kilonova lightcurve are the overall luminosity and the relative colors in the photometric bands. The former is related to the amount of matter as well as the object's distance, while the latter is related to its composition, such as the lanthanide fraction, and viewing angle to the binary. The mass of the unbound material ejected from the system is a key parameter for the computation of kilonova lightcurves. The ejecta mass and further ejecta properties depend on both the nature of the binary -- a BNS, NSBH, or BBH -- and the supra-nuclear equation of state describing the neutron star material. 

\label{Dynamical Ejecta}
\subsection{Equation of state of neutron star}
Low latency / near real-time GW searches do not provide information concerning the compactness of the compact objects. But, for a fixed equation of state that does not support twin stars, fixing the mass of a neutron star fixes the baryonic mass $m^{\rm bar}$. Equally, it fixes the radius $R$ and also the compactness by means of the relation $C = \frac{Gm}{Rc^2}$, where $G$, $m$, and $c$ are the gravitational constant, the mass of the compact object, and the speed of light in vacuum.
In the literature, there are several equation of state candidates. One possibility is to assume a popular one, e.g.~\citet{Douchin:2001sv}, or to sample a number of equations of state simultaneously, e.g.,~\citet{PhysRevD.99.084049,Capano:2019eae,Dietrich:2020lps}.
We do the latter using the 4-parameter spectral representation of the equation of state presented in \citet{Abbott:2018exr}. More specifically, the spectral representation decomposes the equation of state's adiabatic index $\Gamma$ into a polynomial in the logarithm of the pressure with coefficients $\{\gamma\} = (\gamma_0, \gamma_1, \gamma_2, \gamma_3)$~\citep{Lindblom:2010bb, 2012PhRvD..86h4003L, Lindblom:2013kra}. Given the specification of a low-density crust model, which we take to be \textit{Skyrme Lyon} (SLY)~\citep{Douchin:2001sv}, and the requirement of smooth matching, the equation of state is uniquely specified by its spectral parameters.

For every $(m_1, m_2)$, we sample independently the compactness for each component. To do so, we marginalize over the 2396 GW170817-like equations of state presented in \cite{Abbott:2018exr}.
For each equation of state, the compactnesses $C_1=C(m_1)$ and $C_2=C(m_2)$, as well as the baryonic mass of the lighter object $m^{\rm bar}_2=m^{\rm bar}(m_2)$ are calculated, and  a maximum neutron star mass is prescribed. 
For each sample, if one of the components has a mass higher than this threshold (defined by the equation-of-state dependent maximum neutron star mass), it is considered to be a black hole.\footnote{Note that spinning NSs can support $\sim20\%$ more mass~\citep{10.1093/mnras/stw575}, but we neglect this as all known Galactic NS have relatively low spins.}
This allows us to put each sample in one of the three categories: BNS, NSBH, and BBH. This marginalization procedure yields a list of  7-tuples $(q, m_\mathrm{chirp}, \chi_\mathrm{eff}, C_1, C_2, m^{\rm bar}_2, f)$, where $f \in \{0, 1, 2\}$ stands for the type of binary: BNS ($f=0$) or NSBH ($f=1$) or BBH ($f=2$). The size of this list of samples is equal to the number of initial MBTA templates times 2396 (the number of equations of states). For those samples consistent with being BBHs, we assume that there are no ejecta. For the BNS and NSBH cases, we calculate the ejecta mass and velocity as described in the following. 



\subsection{Ejecta parameters: BNS}
In general, there are (at least) two ejecta mass components contributing to the kilonova: the dynamical ejecta and the disc mass.
We follow \citet{Dietrich:2020lps} and use the formula $m_\mathrm{ej} = M_{\rm ej}^{\rm dyn} + \zeta M_{\rm disc}$ to represent the ejecta proportions.
Here, $M_{\rm ej}^{\rm dyn}$ stands for the mass of the dynamical ejecta, and $M_{\rm disc}$ stands for the disc mass. From the disc a fraction of matter ($\zeta$) will become unbound through disc winds caused by several physical phenomena, e.g., neutrino radiation, magnetic-driven winds, or the redistribution of angular momentum. We assume, based on numerical-relativity simulations, that $\zeta = 0.15$ of the entire disc mass gets gravitationally unbound and ejected from the system, e.g.,~\citep{Fernandez:2014cna,Siegel:2017jug,Fernandez:2018kax,Christie:2019lim}.

To estimate the dynamical ejecta, we use the fitting formula from~\citet{Coughlin:2018fis}:
\[\log_{10}M_\mathrm{ej}^\mathrm{dyn}(M_\odot) = \left[a\frac{(1 - 2 C_1)m_1}{C_1} + b\text{ }m_2\left(\frac{m_1}{m_2}\right)^n + \frac{d}{2}\right] + \left[1 \leftrightarrow 2\right],\]
where $m_1$ and $C_1$ (respectively $m_2$ and $C_2$) are the mass and the compactness of the heavier (respectively lighter) binary component, and $a= -0.0719$, $b=0.2116$, $d=-2.42$, $n = -2.905$ are fitting coefficients. 
To estimate the disc mass, we use the fitting formula from~\citet{Dietrich:2020lps}:
\begin{small}
\[\log_{10}M_{\rm disc}(M_\odot) = \max \left(-3, a\left(1 + b\tanh\left(\frac{c - (m_1 + m_2)/M_{\rm thresh}}{d}\right)\right) \right), \]
\end{small}
the floor value of $10^{-3} M_\odot$ is added as it is difficult to resolve smaller masses in numerical relativity (see, e.g., \cite{DiUj2017,Radice:2018pdn}).
Here, $M_{\rm thresh}$ is the minimum total mass such that the prompt collapse occurs after the coalescence of the two neutron stars; this expression is calculated as in~\cite{Bauswein:2013jpa}. While the parameters $c=0.953$ and $d=0.0417$ are fixed, the parameters $a$ and $b$ are not constant but mass ratio dependent; cf.~\cite{Dietrich:2020lps} for a detailed discussion. 

\begin{figure*}
    \includegraphics[width=3.in]{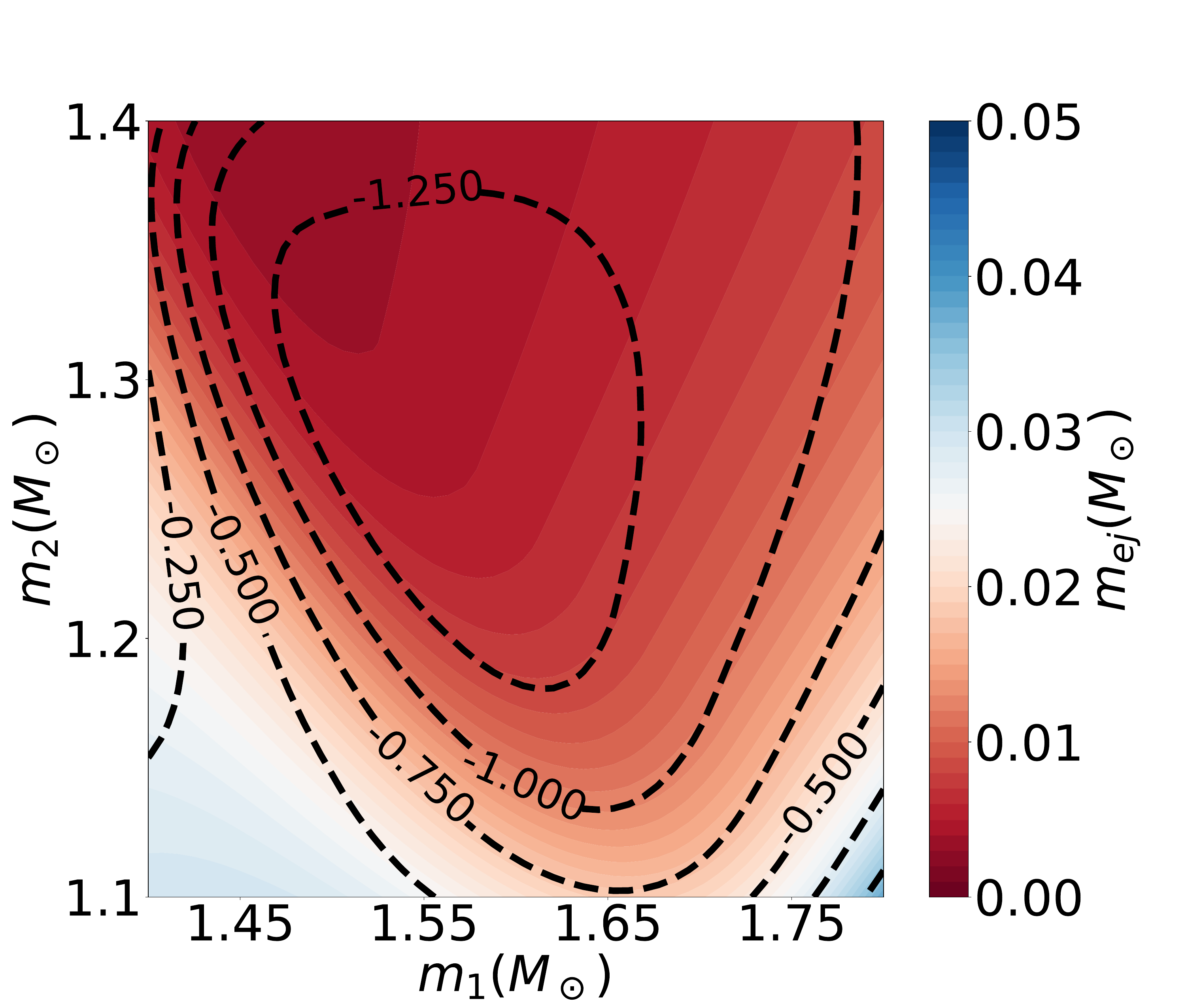}
    \includegraphics[width=3.in]{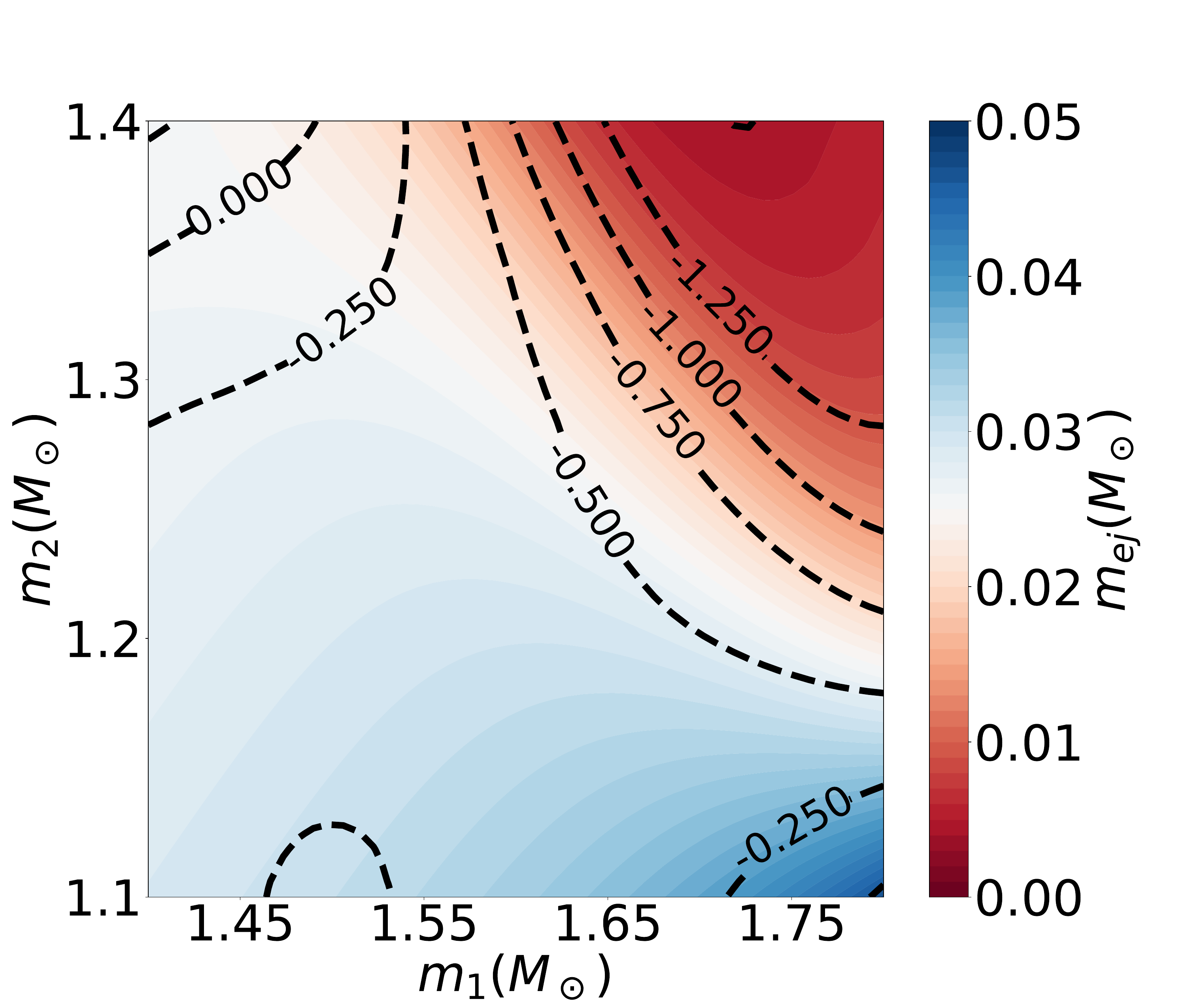}
    \caption{
    In color we show the predicted mass ejecta obtained by our model for BNS mergers with different binary component masses. The level lines show the relative difference which exists between our model and the one from~\citep{Nedora:2020qtd}. The quantity illustrated by those lines is $2 ({m_{\rm ej} - m_{\rm ej}^{\rm NEA}})/({m_{\rm ej} + m_{\rm ej}^{\rm NEA}})$, where $m_{\rm ej}^{\rm NEA}$ is the predicted mass ejecta computed by the model in~\citep{Nedora:2020qtd}.
    The left (respectively right) panel corresponds to a fixed spectral equation of state with parameters $\{\gamma\} = (0.5485,0.3767,-0.0690,0.0035)$ (respectively $\{\gamma\} = (1.4777, -0.3225, 0.0694, -0.0046)$); this equation of state predicts a radius of $11.3$ km ($13.0$ km) and a tidal deformability of $232$ ($663$) for a $1.4M_\odot$ neutron star, as well as a maximum Tolman-Oppenheimer-Volkov mass of $2.00M_\odot$ ($2.43M_\odot$).
    }
    \label{fig:bns_ejected}
\end{figure*}

For comparison, in~\cite{Nedora:2020qtd}, the disc and dynamical masses are calculated by means of a formula using mass ratio and the tidal deformability $\tilde\Lambda$. An illustration of the ejecta dependence on the binary component masses, as well as a comparison to the predictions of this latter model, is illustrated in Figure~\ref{fig:bns_ejected}.
This demonstrates broad qualitative consistency, but differences of $\sim$\,100\% between different predictions are common.
This is mainly due to different sets of numerical-relativity simulations that are used for the calibration and different functional forms for the phenomenological fits. In addition, one has to point out that while the numerical-relativity simulations provide a description of the merger and postmerger dynamics and are capable of predicting (to some extent) the amount of ejecta, the individual predictions are usually connected to large uncertainties due to, among others, (i) the absence of an accurate microphysical modelling of the fluid as well as the inclusion of magnetic fields, (ii) the complications during the simulation of the relativistic fluids, when shocks and discontinuities form, (iii) inaccuracies during the simulation of the expanding and decompressing ejected material, and (iv) a limited set of numerical simulation that do not cover the entire BNS parameter space.
Nonetheless, some relations between the binary parameters and the amount of ejecta are noticeable and we find that lower compactness and/or smaller individual masses produce in general more ejecta.

We also compute the velocity of the ejecta. Following \citet{Coughlin:2018fis}, we use $v_{\rm ej} = \left[a(1 + c\text{ }C_1\frac{m_1}{m_2} + \frac{b}{2})\right] + [1 \leftrightarrow 2]$, where fit coefficients are $a=-0.3090$, $b=0.657$ and $c=-1.879$. The result in this formula is expressed in units of the speed of light.

\subsection{Ejecta parameters: NSBH}
Similar to the BNS merger case, we assume that NSBH ejecta have (at least) two components: the dynamical ejecta and disc wind ejecta. In a NSBH system, the only baryonic matter responsible for any EM signature is the one contained in the neutron star, i.e., $m^{\rm bar}_2$. 
As in the BNS case, we assume that 15\% of the disc mass becomes gravitationally unbound over time, where the disc mass is estimated according to~\citet{Foucart:2018rjc} as 
\[M_{\rm disc}(M_\odot) = m_2^{\rm bar} \max\left(0, \alpha\frac{1 - 2C_2}{\eta^{1/3}} - \beta r_{\rm ISCO} \frac{C_2}{\eta} + \gamma\right)^\delta,\]
with $m_2^{\rm bar}$, $C_2$, $\eta = \frac{m_1 m_2}{m_1 + m_2}$, and $r_\mathrm{ISCO}$ being the baryonic mass of the neutron star, the compactness of the neutron star, the reduced mass, and the innermost stable circular orbit. The coefficients are $\alpha = 0.4064$, $\beta = 0.1388$, $\gamma = 0.2551$, and $\delta = 1.7612$.  The mass of the dynamical ejecta is calculated from~\cite{Kruger:2020gig}
\[M^{\rm dyn}(M_\odot) = m_2^{\rm bar}\left(a_1 \left(\frac{m_1}{m_2}\right)^{n_1} \frac{1 - 2C_2}{C_2} - a_2\left(\frac{m_1}{m_2}\right)^{n_2}\frac{r_{\rm ISCO}}{m_1} + a_4\right).\]
In this formula, $m_1$ is the mass of the black hole and the fitting coefficients are $a_1 = 0.007116 $, $a_2 = 0.001436$, $a_4 = -0.02762$, $n_1 = 0.8636$ and $n_2 = 1.6840$.

\begin{figure*}
    \includegraphics[width=2.25in]{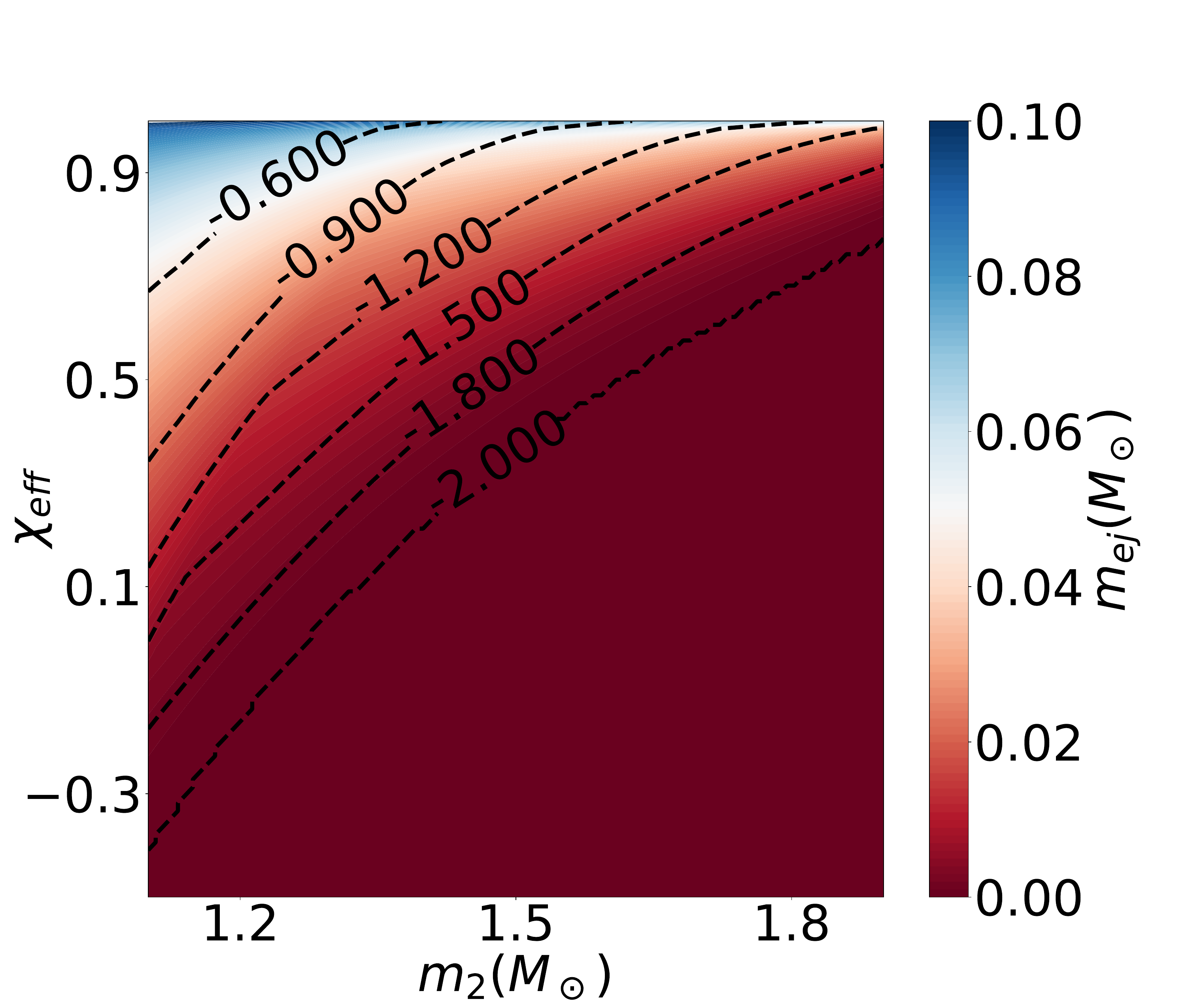}
    \includegraphics[width=2.25in]{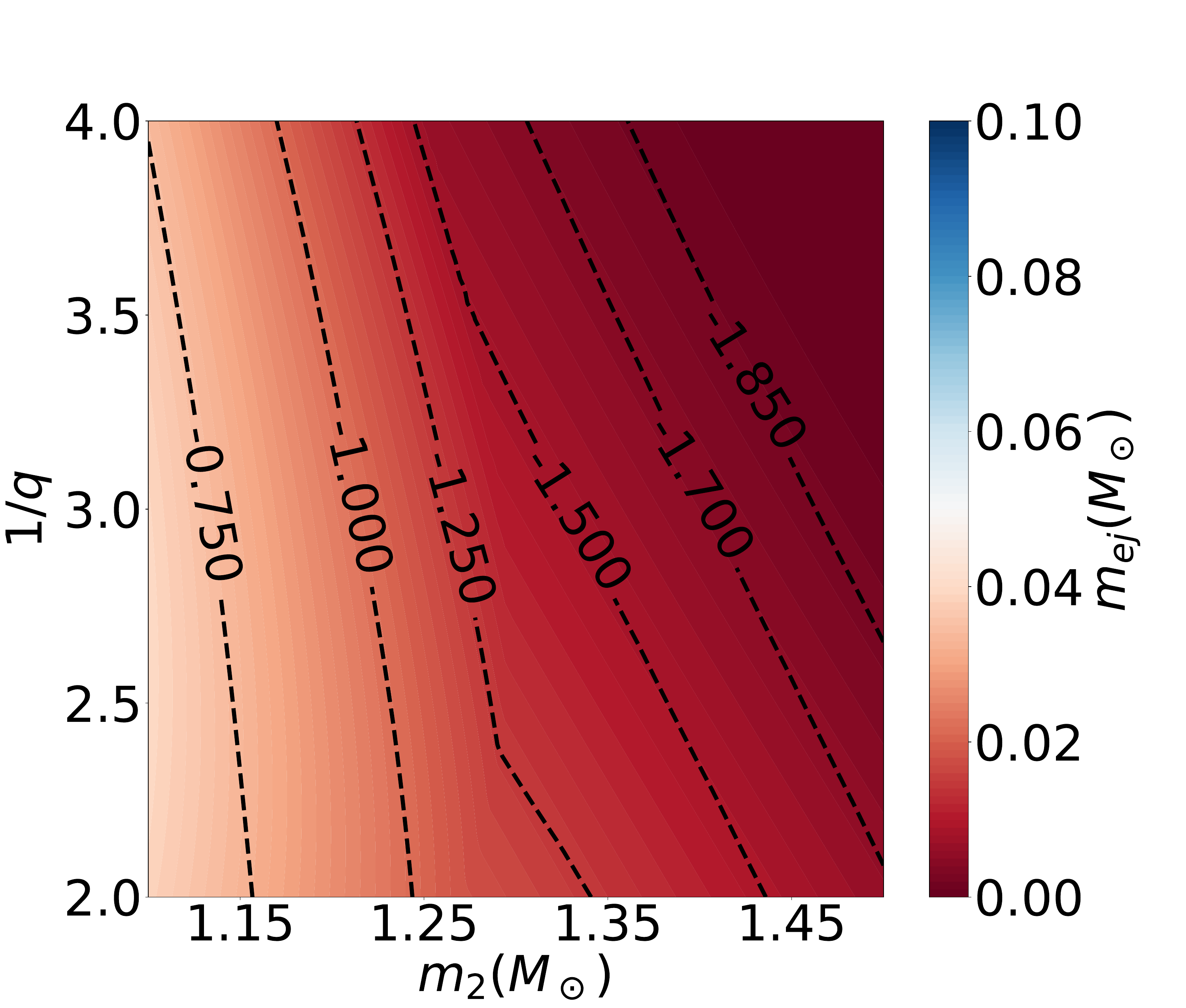}
    \includegraphics[width=2.25in]{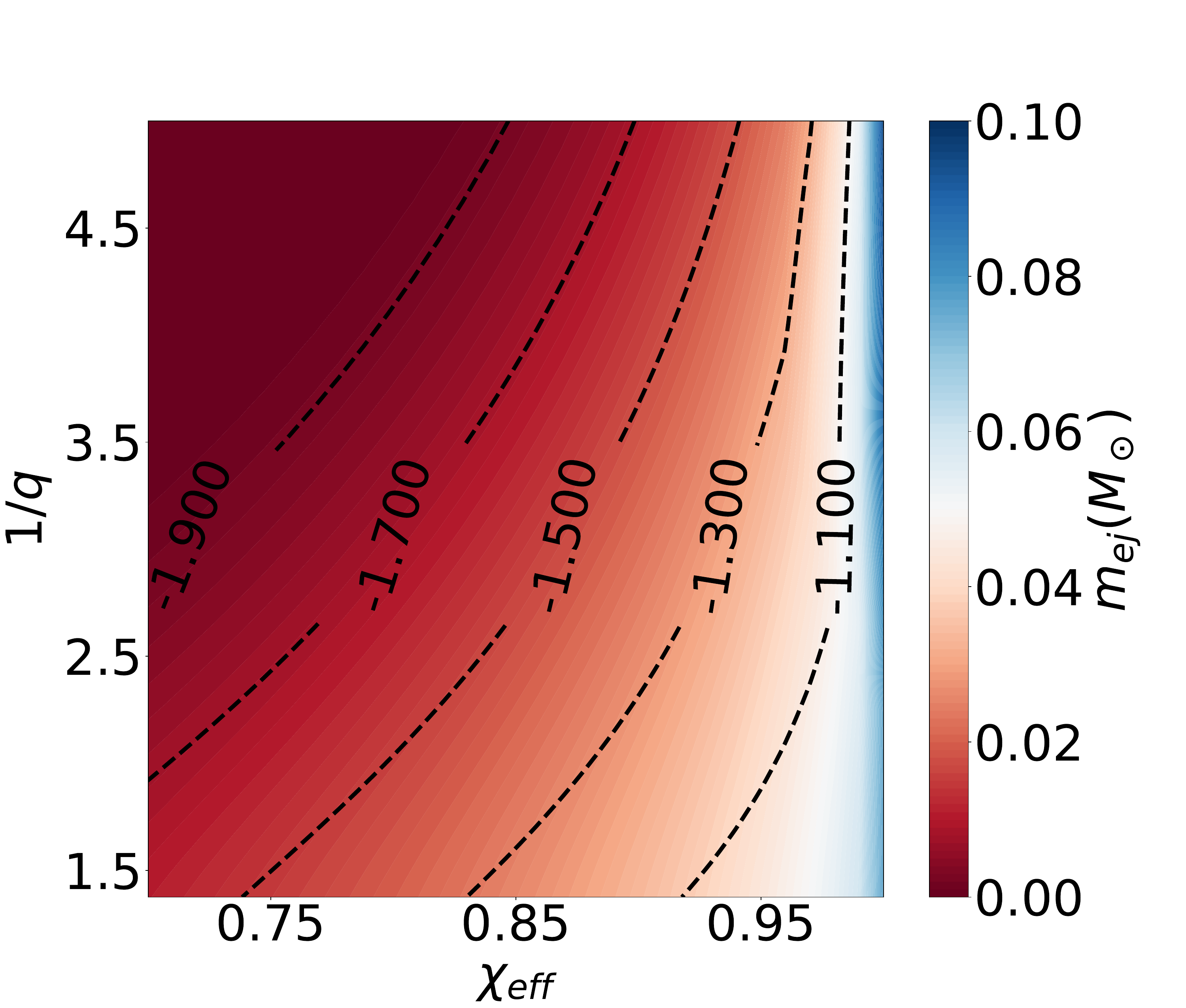}
    \caption{Amount of ejected mass (in color axis) for NSBH mergers with different binary characteristics. $\chi_\mathrm{eff}$ versus $m_2$, keeping $1/ q =m_1 / m_2 =2$ (left panel); inverse mass ratio $1/q$ versus neutron star mass $m_2$, keeping $\chi_\mathrm{eff} = 0.6$ (middle panel); and $1/q$ versus $\chi_\mathrm{eff}$ at constant $m_2 = 1.6 M_{\odot}$ (right panel). The level lines show the dynamical ejecta relative difference which exists between our model and the one proposed in~\citep{Kawaguchi:2016ana}. More precisely, the values of the level lines indicate the quantity $2(M^{\rm dyn} - M^{\rm dyn}_{\rm KAW}) / (M^{\rm dyn} + M^{\rm dyn}_{\rm KAW})$, where $M^{\rm dyn}_{\rm KAW}$ is the dynamical ejecta predicted by~\citep{Kawaguchi:2016ana}. These simulations use the equation of state with parameters $\{\gamma\} = (0.3268,0.4456,-0.0586,0.0016)$, which predicts a radius of $12.4$ km and a tidal deformability of $458$ for a $1.4M_\odot$ neutron star, the maximum Tolman-Oppenheimer-Volkov mass being $2.37M_\odot$.}
    \label{fig:nsbh_ejected}
\end{figure*}

The dependence of mass ejected with the various parameters is illustrated in Figure~\ref{fig:nsbh_ejected}. One can easily observe that the higher the effective spin, the higher the mass ejecta. At very high inverse mass ratios, for a constant $m_2$, the neutron star is swallowed by the black hole before being disrupted; cf.~\cite{Shibata:2011jka,Foucart:2020ats} and references therein for a detailed description.~\cite{Kawaguchi:2016ana} proposes similar ejecta fits for the dynamical ejecta. A comparison between those predicted dynamical ejecta and our choice of $M^{\rm dyn}$ is equally proposed in Figure~\ref{fig:nsbh_ejected}.

We also compute the velocity of the ejecta. Following \cite{Kawaguchi:2016ana}: $v_{\rm ej} = \alpha\frac{m_1}{m_2} + \beta$ with $\alpha = 0.01533$ and $\beta = 0.1907$. The result in this formula is expressed in units of the speed of light.

\section{Lightcurves}
\label{sec:surrogates}
Once the calculations presented in the previous section are done we have 2396 times $n_{\rm templates}$ $(m_\mathrm{ej}, v_\mathrm{ej})$ tuples, where $n_{\rm templates}$ stands for the number of MBTA templates. This set is downsampled to a set of size 1,000 for computational cost reasons. 
The value 1,000 is justified by the similarity between the $m_\mathrm{ej}$ distributions of the initial and the downsampled set. 
Such a size of the downsampled set allows an overlap higher than 80\% between the initial and the new mass ejecta distributions in the case of GW170817 and GW190425. 
We now use lightcurve models to translate the ejecta properties into observed lightcurves. We use the lightcurve models proposed in~\citet{KaMe2017} (hereafter \textit{Model I}) and  \citet{Bulla:2019muo} (hereafter \textit{Model II}). These are radiative transfer simulations predicting lightcurves and spectra, based on the wavelength-dependent emissivity, and opacity taking place at the atomic scale. In particular, we use the surrogate technique first presented in~\citet{Coughlin:2018miv} to create a grid of lightcurves in the photometric bands $u$, $g$, $r$, $i$, $z$, $y$, $J$, $H$, and $K$ for a set of model parameters. Using Gaussian Process Regression, one can predict the lightcurve for any input parameters. A common parameter for the two models is the total ejecta mass, i.e., the sum of the dynamical and wind ejecta. That is equivalent to saying that we use the 1-component ejecta model presented in~\cite{Coughlin:2018miv}, i.e., the EM signal luminosity is calculated at once for the entire ejected matter. This is contrary to the case of 2-component model where the brightness due to disc winds and dynamical ejecta are calculated separately and added up a posteriori. 
Therefore the statistical errors regarding the 2-component model are already large enough that this choice does not make a difference; 
cf.~e.g.~\citep{Kawaguchi:2019nju,Heinzel:2020qlt} for a discussion about uncertainties and viewing angle dependencies of the kilonova signal. 
In addition, for \textit{Model II} we use the grid first presented in~\cite{2020NatCo..11.4129C}, but extended to better cover the lower- and upper-mass end (it goes from $10^{-6}M_\odot$ to $1M_\odot$). This upgrade will be made publicly available.\footnote{\url{https://github.com/mbulla/kilonova_models }} It is noteworthy to mention that for all the lightcurve contours presented in this section, two extra magnitude errors have been added, i.e. the upper (lower) magnitude limits have been raised (lowered) by 1, in order to be robust against twice the errors in thermalization rate and/or ejecta geometry.   

\subsection{\textit{Model I} lightcurve model}
\textit{Model I} presented in~\cite{KaMe2017} solves the relativistic radiation transport Boltzmann equation governing the interior of a radioactive plasma. In this way, both the thermal and spectral-lines radiation determine the final wavelength-dependent luminosity and timescale of the lightcurve.
\textit{Model I} lightcurve is a function of $m_{\rm ej}, v_{\rm ej}$ and $X_{\rm lan}$, where $m_{\rm ej}$ is the ejected mass, $v_{\rm ej}$ is the velocity of the ejecta, and $X_{\rm lan}$ is the lanthanide fraction. The effects of the first two parameters are simple and intuitive: the higher the amount of ejecta, the brighter and longer-lasting is the electromagnitic signal; meanwhile the higher the speed of the ejecta; the brighter and shorter (the ejecta is expanding faster) is the kilonova. The latter parameter $X_{\rm lan}$ expresses the composition of the ejecta and controls the opacity at the atomic scale. Therefore, for ejecta containing heavier elements, the density of spectral lines is larger. This aspect will imply a higher opacity and with that a fading of the brightness on larger timescales.

\begin{figure}
    \includegraphics[width=3.in]{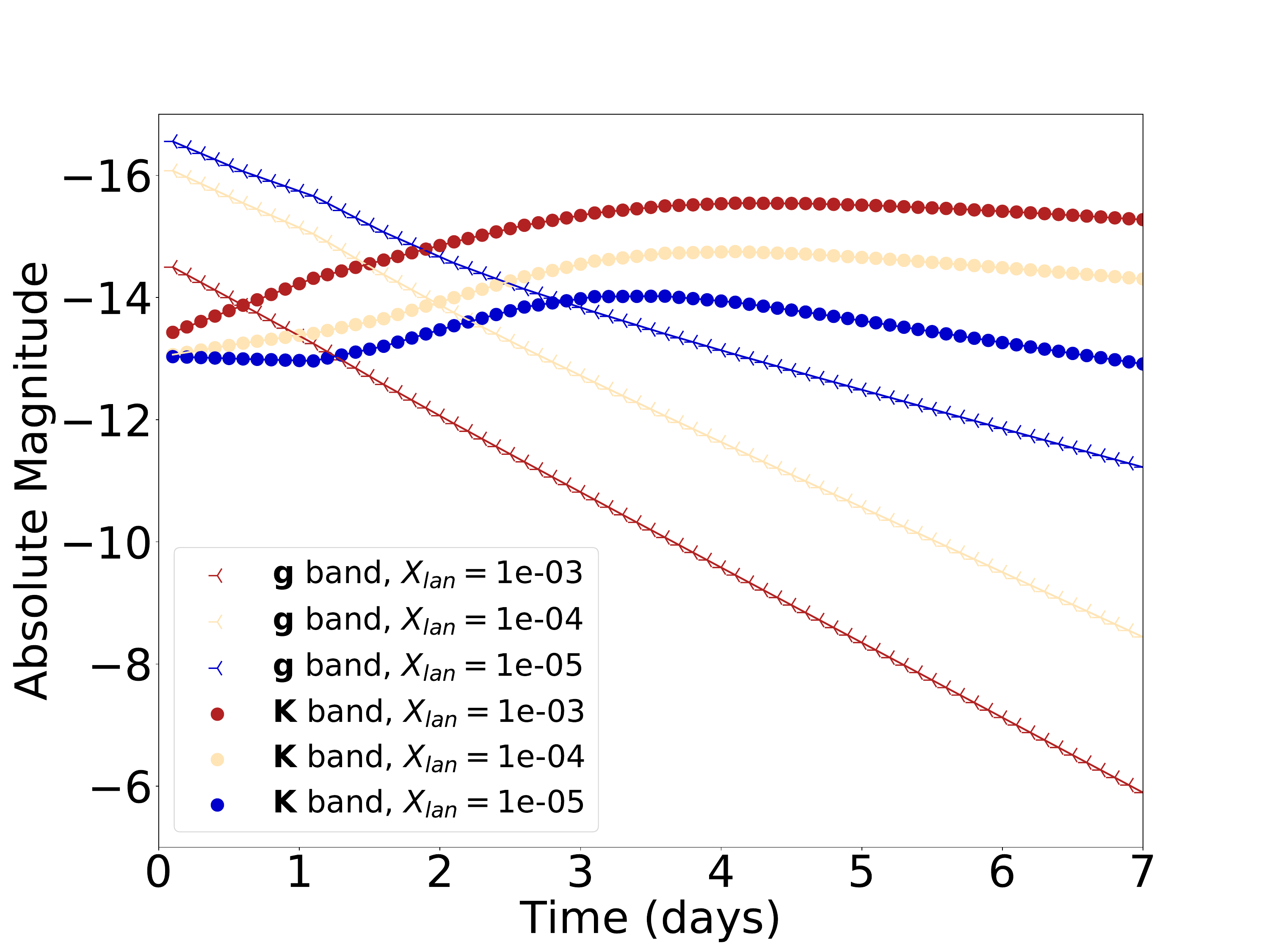}
    \caption{Absolute magnitude versus time for kilonovae with different lanthanide percentage $X_\mathrm{lan}$, as predicted by \textit{{Model I}}. For all the lightcurves the values of mass ejecta and velocity of ejecta are set to $m_\mathrm{ej} = 0.05 M_\odot$ and $v_\mathrm{ej} = 0.15c$. The plot corresponds to the 'g' and 'K' photometric bands. The x axis origin corresponds to the merger time.
    }
    \label{fig:Kasen_Xlan}
\end{figure}

We compute $m_{\rm ej}$ and $v_{\rm ej}$ as described in Section~\ref{Dynamical Ejecta}; $X_{\rm lan}$, on the other hand, requires further assumptions. In general, a larger $X_{\rm lan}$ yield redder lightcurves. In Figure~\ref{fig:Kasen_Xlan}, there is an example of the dependence of lightcurve with lanthanide fraction. One can observe that an uncertainty in $X_{\rm lan}$ of 3 orders of magnitude leads to an uncertainty in the 'g' band of more than 2 (respectively 5) magnitudes at the end of the  first (seventh) day .

\subsection{\textit{Model II} lightcurve model}
\textit{Model II}~\citep{Bulla:2019muo, 2020NatCo..11.4129C} is also based on Monte-Carlo radiative transfer simulations. Unlike \textit{Model I}, which is 1D and has geometry-independent parameters, \textit{Model II} is 2D, i.e., axisymmetric, and the ejecta are considered to have two components with different compositions and whose locations are determined with respect to the geometry of the binary. Therefore, the EM signal depends on the position of the external observer, i.e., the viewing angle. In this model, one component is lanthanide-rich and is situated around the plane of the merger, whereas a second component is lanthanide-free and positioned at higher latitudes. The interplay between the two components is captured by the half opening angle of the lanthanide-rich component, $\Phi$, while the position of the observer is controlled by the viewing angle $\theta_{\rm inc}$. 

The two models differ in their considerations of ejecta opacities. While in \textit{Model I} the lanthanide fraction can take whatever value and the opacities are calculated correspondingly, in \textit{Model II} the composition of the two ejecta components are fixed and for simplicity we assume just two different compositions and corresponding opacities (one for each component, see \citep{Bulla:2019muo}).

\begin{figure}
    \includegraphics[width=3.in]{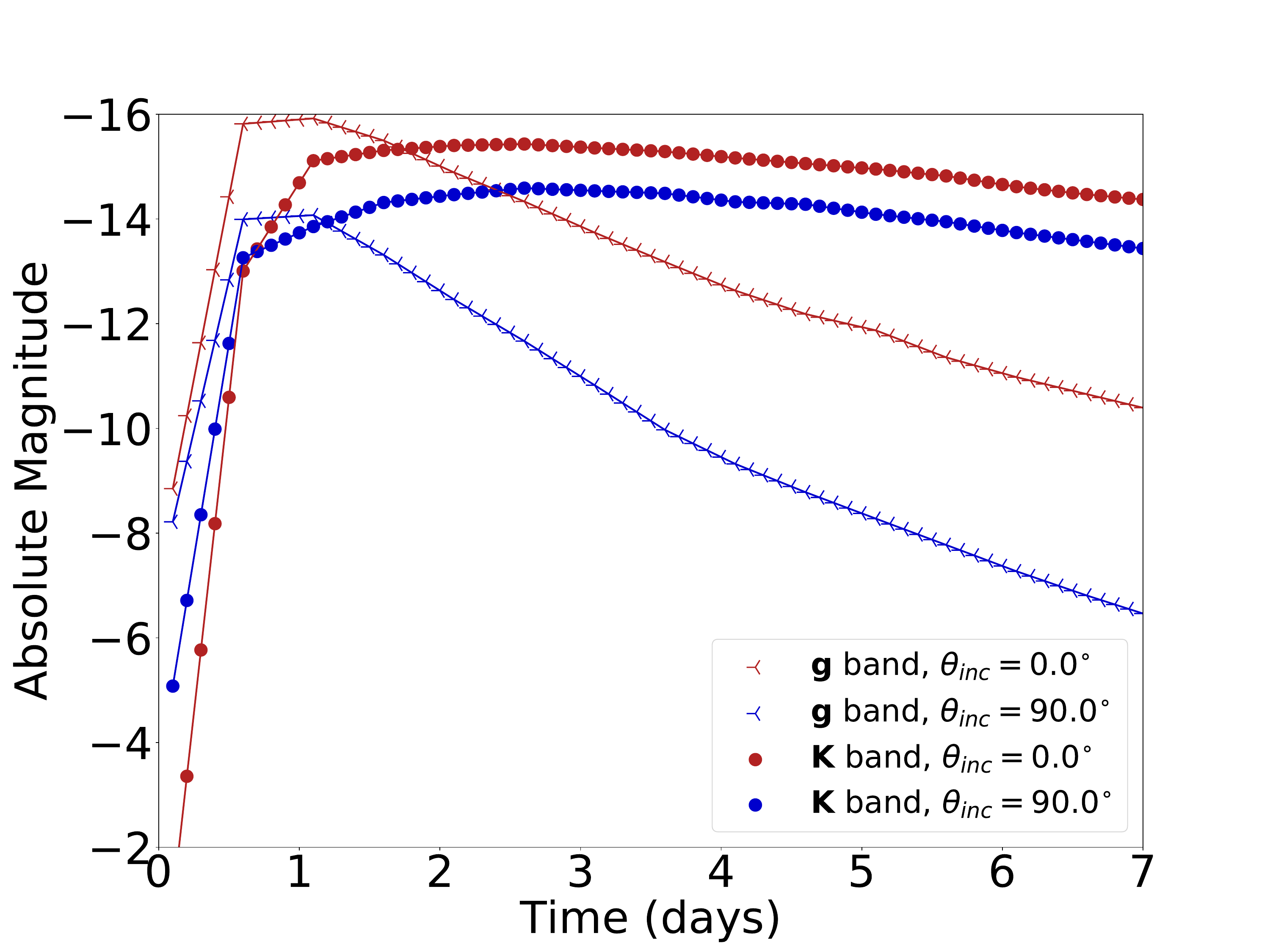}
    \includegraphics[width=3.in]{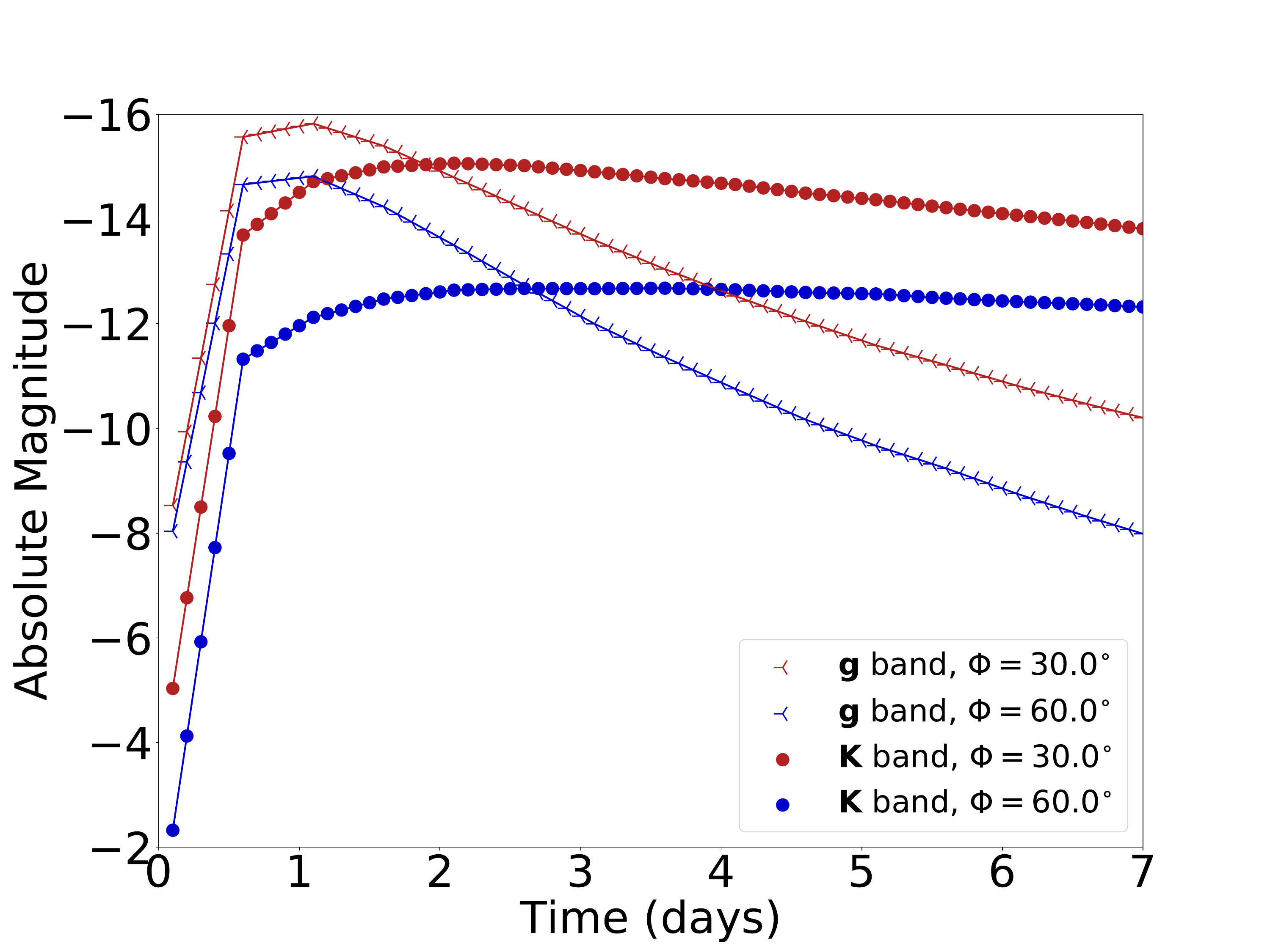}
    \caption{Absolute magnitude versus time for kilonovae with different inclination angles $\theta_\mathrm{inc}$ (top) and opening angles $\Phi$ (bottom) as expected in 'g' and 'K' photometric passband and as predicted by \textit{Model II}. For the top plot, we set $m_\mathrm{ej}=0.05 M_\odot$ and $\Phi=45^\circ$, while for the bottom plot we set $m_\mathrm{ej}=0.05 M_\odot$ and $\theta_\mathrm{inc}=45^\circ$. On both sides, the x axis origin corresponds to the merger time.}
    \label{fig:Bulla_theta_phi}
\end{figure}

\textit{Model II} depends on the following parameters: $m_{\rm ej}$, $\theta_{\rm inc}$, and $\Phi$, where $m_{\rm ej}$ is as above the total mass of the ejecta. Because MBTA samples do not provide $\theta_{\rm inc}$ for O2 and O3, and there is no imprint of $\Phi$ in the GW signal, some additional assumptions are necessary. For illustration purposes, Figure~\ref{fig:Bulla_theta_phi} shows the dependence of the EM lightcurves for different prior choices, where an increase in the opening angle reddens the lightcurve, whereas an increase in the inclination angle will lower the luminosity and redden the signal.

\subsection{Sources of uncertainty}
\label{sec:uncertainties}
Our predicted lightcurves have assigned uncertainties. These are due to either the inaccurate measurement of the GW strain by the GW interferometers or our limited knowledge about the composition of the stars and the way matter behaves at supra-nuclear densities. More specifically, there are the following sources of uncertainty: the inaccurate measurement of the binary parameters such as the chirp mass, the mass ratio and the dimensionless effective spin; the uncertainty induced by the GW170817-like equation of state marginalization; the errors produced by the mass (as well as the velocity) ejecta  fits; and finally the missing knowledge about the ejecta chemical composition, which in the case of \textit{Model I} (respectively \textit{Model II}) is represented by the lanthanide fraction (respectively the half opening angle of the lanthanide-rich ejecta component). In this section we illustrate the impact of these uncertainty sources on the lightcurves and on \textit{HasEjecta}, defined as the probability of having $m_{\rm ej} > 3\times10^{-4}M_\odot$. The value of this threshold is argued by the fact that $1.5 \times 10^{-4} M_\odot$ is the minimum mass ejecta for a BNS (based on the disc wind mass fit), and as a consequence we consider that a configuration produces noticeable ejecta when the total ejecta is at least twice as large as this default value. We start with a binary whose parameters are fixed, $m_{\rm chirp}^{\rm fixed}, q^{\rm fixed}, \chi_{\rm eff}^{\rm fixed}$. Moreover, we assume a fixed neutron star equation of state, ${\rm EOS}^{\rm fixed}$. If in addition we assume that the ejecta fits have no errors and we fix the lanthanide fraction to $X_{\rm lan}^{\rm fixed}$, then the \textit{Model I} lightcurves have no uncertainty. From Table~\ref{tab:table_limits}, one can see that the low-latency pipelines constrain well the chirp mass (around 1\% error), while the measured mass ratio and the effective spin have large errors, sometimes overcoming 100\%. Wherefore, in order to assess the effect of the low-latency pipelines inaccurate measurements, we consider the uniform grid points $(m_{\rm chirp}, q, \chi_{\rm eff}) \in [0.99 m_{\rm chirp}^{\rm fixed}, 1.01 m_{\rm chirp}^{\rm fixed}] \times [\frac{1}{2}q^{\rm fixed}, \min{(2 q^{\rm fixed}, 1)}] \times [-\chi_{\rm eff}^{\rm fixed}, \min{(3 \chi_{\rm eff}^{\rm fixed}, 1)}]$. On all our examples $\chi_{\rm eff}^{\rm fixed} \geq 0$. On the other hand, the equation of state ${\rm EOS}^{\rm fixed}$ and $X_{\rm lan}^{\rm fixed}$ are unchanged and we always assume that the ejecta fits have no errors. Similarly, the impact of equation of state marginalization on the lightcurves output is derived by considering the entire set of 2396 equations of state and keeping all the other parameters set to their initial fixed values. Figures~\ref{fig:bns_ejected} and~\ref{fig:nsbh_ejected} show that by using different ejecta fits, one may obtain noticeably different values for the merger expelled matter. The effect of the mass and velocity ejecta fits are probed by considering uniform grid points $(m_{\rm ej}, v_{\rm ej}) \in [\frac{1}{3}m_{\rm ej}^{\rm fixed}, 3 m_{\rm ej}^{\rm fixed}] \times [\frac{1}{2}v_{\rm ej}^{\rm fixed}, 2 v_{\rm ej}^{\rm fixed}]$ and keeping the lanthanide fraction equal to $X_{\rm lan}^{\rm fixed}$. In the preceding expression, $m_{\rm ej}^{\rm fixed}, v_{\rm ej}^{\rm fixed}$ are the mass and velocity of ejecta obtained from $m_{\rm chirp}^{\rm fixed}$, $q^{\rm fixed}$ and $\chi_{\rm eff}^{\rm fixed}$. Equally, our limited knowledge about the composition of the ejecta is evaluated by considering one dimensional uniform grid points $\log_{10}{X_{\rm lan}} \in [-9, -1]$, at fixed $m_{\rm ej} = m_{\rm ej}^{\rm fixed}$ and $v_{\rm ej} = v_{\rm ej}^{\rm fixed}$. Finally, we also treat the case of all uncertainty sources combined: we start with $(m_{\rm chirp}, q, \chi_{\rm eff}) \in [0.99 m_{\rm chirp}^{\rm fixed}, 1.01 m_{\rm chirp}^{\rm fixed}] \times [\frac{1}{2}q^{\rm fixed}, \min{(2 q^{\rm fixed}, 1)}] \times [-\chi_{\rm eff}^{\rm fixed}, \min{(3 \chi_{\rm eff}^{\rm fixed}, 1)}]$; then we marginalize over the entire set of 2396 equations of state; for each sample $(m_{\rm ej}^{\rm predicted}, v_{\rm ej}^{\rm predicted})$ of predicted mass and velocity ejecta we consider a distribution of samples in $[\frac{1}{3}m_{\rm ej}^{\rm predicted}, 3 m_{\rm ej}^{\rm predicted}] \times [\frac{1}{2}v_{\rm ej}^{\rm predicted}, 2 v_{\rm ej}^{\rm predicted}]$; finally we marginalize with $\log_{10}(X_{\rm lan})$ uniformly sampled in $[-9, -1]$. 
\begin{table*}
    \centering
    \begin{tabular}{|c|c|c|c|c|c|c|c|c|c|c|}
     \hline 
     \cline{6-11}
     \multicolumn{3}{|c}{Binary} & \multicolumn{1}{|c}{}& \multicolumn{1}{|c}{}& \multicolumn{6}{|c|}{Absolute magnitude} \\
     \cline{1-3} \cline{6-11}
     \multicolumn{1}{|c}{$m_1^{\rm fixed}$} & \multicolumn{1}{|c}{$m_2^{\rm fixed}$} & \multicolumn{1}{|c}{$\chi_{\rm eff}^{\rm fixed}$} & \multicolumn{1}{|c}{} &
     \multicolumn{1}{|c}{\textit{HasEjecta}} &
     \multicolumn{2}{|c}{1 day} & \multicolumn{2}{|c}{2 days} & \multicolumn{2}{|c|}{3 days}\\

     \multicolumn{1}{|c}{$(M_\odot)$} & \multicolumn{1}{|c}{$(M_\odot)$} & \multicolumn{1}{|c}{} & \multicolumn{1}{|c}{Source of uncertainty} &
     \multicolumn{1}{|c}{\textit{(\%)}} &
     \multicolumn{1}{|c}{$\mathit{\mathbf{g}}$ band} & \multicolumn{1}{|c}{$\mathit{\mathbf{K}}$ band} &  \multicolumn{1}{|c}{$\mathit{\mathbf{g}}$ band} & \multicolumn{1}{|c}{$\mathit{\mathbf{K}}$ band} &  \multicolumn{1}{|c}{$\mathit{\mathbf{g}}$ band} & \multicolumn{1}{|c|}{$\mathit{\mathbf{K}}$ band}\\
    \hline
    \multicolumn{1}{|c}{} & \multicolumn{1}{|c}{} & \multicolumn{1}{|c}{} & \multicolumn{1}{|c}{no uncertainty} & \multicolumn{1}{|c}{} & \multicolumn{1}{|c}{-13.5} & \multicolumn{1}{|c}{-12.7} & \multicolumn{1}{|c}{-11.9} & \multicolumn{1}{|c}{-13.0} & \multicolumn{1}{|c}{-10.3} & \multicolumn{1}{|c|}{-12.6} \\ \multicolumn{1}{|c}{} & \multicolumn{1}{|c}{} & \multicolumn{1}{|c}{} & \multicolumn{1}{|c}{MBTA} & \multicolumn{1}{|c}{100} & \multicolumn{1}{|c}{$-13.7^{-1.9}_{+0.5}$} & \multicolumn{1}{|c}{$-12.9^{-1.1}_{+0.4}$} & \multicolumn{1}{|c}{$-12.2^{-2.3}_{+0.7}$} & \multicolumn{1}{|c}{$-13.4^{-1.1}_{+0.7}$} & \multicolumn{1}{|c}{$-10.7^{-2.8}_{+0.8}$} & \multicolumn{1}{|c|}{$-13.1^{-1.9}_{+1.0}$} \\ \multicolumn{1}{|c}{1.6} & \multicolumn{1}{|c}{1.4} & \multicolumn{1}{|c}{0.01} & \multicolumn{1}{|c}{equation of state} & \multicolumn{1}{|c}{100} & \multicolumn{1}{|c}{$-11.9^{-1.6}_{+0.3}$} & \multicolumn{1}{|c}{$-12.1^{-0.6}_{+0.2}$} & \multicolumn{1}{|c}{$-10.0^{-1.9}_{+0.6}$} & \multicolumn{1}{|c}{$-11.6^{-1.5}_{+0.3}$} & \multicolumn{1}{|c}{$-8.2^{-2.2}_{+0.8}$} & \multicolumn{1}{|c|}{$-11.2^{-1.4}_{+0.2}$} \\ \multicolumn{1}{|c}{} & \multicolumn{1}{|c}{} & \multicolumn{1}{|c}{} & \multicolumn{1}{|c}{$m_{\rm ej}$, $v_{\rm ej}$} & \multicolumn{1}{|c}{} & \multicolumn{1}{|c}{$-13.5^{-0.5}_{+1.5}$} & \multicolumn{1}{|c}{$-12.9^{-0.8}_{+1.2}$} & \multicolumn{1}{|c}{$-11.8^{-1.0}_{+2.3}$} & \multicolumn{1}{|c}{$-12.8^{-1.0}_{+0.6}$} & \multicolumn{1}{|c}{$-10.3^{-1.7}_{+3.1}$} & \multicolumn{1}{|c|}{$-12.1^{-0.8}_{+0.3}$} \\ \multicolumn{1}{|c}{} & \multicolumn{1}{|c}{} & \multicolumn{1}{|c}{} & \multicolumn{1}{|c}{$X_{\rm lan}$} & \multicolumn{1}{|c}{} & \multicolumn{1}{|c}{$-14.0^{-0.3}_{+4.8}$} & \multicolumn{1}{|c}{$-12.3^{-2.2}_{+0.3}$} & \multicolumn{1}{|c}{$-13.0^{-0.6}_{+4.9}$} & \multicolumn{1}{|c}{$-12.2^{-2.7}_{+0.9}$} & \multicolumn{1}{|c}{$-12.2^{-1.0}_{+4.9}$} & \multicolumn{1}{|c|}{$-11.3^{-3.4}_{+2.2}$} \\ \multicolumn{1}{|c}{} & \multicolumn{1}{|c}{} & \multicolumn{1}{|c}{} & \multicolumn{1}{|c}{all} & \multicolumn{1}{|c}{100} & \multicolumn{1}{|c}{$-12.9^{-2.8}_{+4.1}$} & \multicolumn{1}{|c}{$-13.0^{-1.7}_{+2.4}$} & \multicolumn{1}{|c}{$-11.9^{-3.0}_{+4.3}$} & \multicolumn{1}{|c}{$-12.9^{-2.1}_{+4.7}$} & \multicolumn{1}{|c}{$-11.1^{-3.3}_{+4.6}$} & \multicolumn{1}{|c|}{$-12.3^{-2.8}_{+6.8}$} \\
    \hline
    \multicolumn{1}{|c}{} & \multicolumn{1}{|c}{} & \multicolumn{1}{|c}{} & \multicolumn{1}{|c}{no uncertainty} & \multicolumn{1}{|c}{} &  \multicolumn{1}{|c}{-13.7} & \multicolumn{1}{|c}{-13.0} & \multicolumn{1}{|c}{-12.1} & \multicolumn{1}{|c}{-13.2} & \multicolumn{1}{|c}{-10.5} & \multicolumn{1}{|c|}{-12.7} \\ \multicolumn{1}{|c}{} & \multicolumn{1}{|c}{} & \multicolumn{1}{|c}{} & \multicolumn{1}{|c}{MBTA} & \multicolumn{1}{|c}{100} & \multicolumn{1}{|c}{$-13.7^{-1.8}_{+2.2}$} & \multicolumn{1}{|c}{$-12.9^{-0.8}_{+1.1}$} & \multicolumn{1}{|c}{$-12.1^{-2.2}_{+2.7}$} & \multicolumn{1}{|c}{$-13.3^{-1.2}_{+2.1}$} & \multicolumn{1}{|c}{$-10.6^{-2.8}_{+3.2}$} & \multicolumn{1}{|c|}{$-13.0^{-2.1}_{+2.1}$} \\ \multicolumn{1}{|c}{2.0} & \multicolumn{1}{|c}{1.4} & \multicolumn{1}{|c}{0.10} & \multicolumn{1}{|c}{equation of state} & \multicolumn{1}{|c}{96} & \multicolumn{1}{|c}{$-13.8^{-0.3}_{+2.9}$} & \multicolumn{1}{|c}{$-13.1^{-0.3}_{+1.8}$} & \multicolumn{1}{|c}{$-12.2^{-0.6}_{+3.0}$} & \multicolumn{1}{|c}{$-13.5^{-0.4}_{+2.5}$} & \multicolumn{1}{|c}{$-10.7^{-0.8}_{+2.6}$} & \multicolumn{1}{|c|}{$-13.0^{-0.3}_{+2.5}$} \\ \multicolumn{1}{|c}{} & \multicolumn{1}{|c}{} & \multicolumn{1}{|c}{} & \multicolumn{1}{|c}{$m_{\rm ej}$, $v_{\rm ej}$} & \multicolumn{1}{|c}{} & \multicolumn{1}{|c}{$-13.6^{-0.3}_{+1.4}$} & \multicolumn{1}{|c}{$-13.2^{-0.6}_{+1.3}$} & \multicolumn{1}{|c}{$-12.1^{-0.8}_{+2.1}$} & \multicolumn{1}{|c}{$-13.0^{-0.7}_{+0.6}$} & \multicolumn{1}{|c}{$-10.7^{-1.4}_{+2.9}$} & \multicolumn{1}{|c|}{$-12.3^{-0.7}_{+0.3}$} \\ \multicolumn{1}{|c}{} & \multicolumn{1}{|c}{} & \multicolumn{1}{|c}{} & \multicolumn{1}{|c}{$X_{\rm lan}$} & \multicolumn{1}{|c}{} & \multicolumn{1}{|c}{$-14.2^{-0.3}_{+4.6}$} & \multicolumn{1}{|c}{$-12.6^{-2.2}_{+0.3}$} & \multicolumn{1}{|c}{$-13.2^{-0.6}_{+4.6}$} & \multicolumn{1}{|c}{$-12.3^{-2.8}_{+1.0}$} & \multicolumn{1}{|c}{$-12.4^{-1.0}_{+4.5}$} & \multicolumn{1}{|c|}{$-11.5^{-3.4}_{+2.3}$} \\ \multicolumn{1}{|c}{} & \multicolumn{1}{|c}{} & \multicolumn{1}{|c}{} & \multicolumn{1}{|c}{all} & \multicolumn{1}{|c}{98} & \multicolumn{1}{|c}{$-12.7^{-2.7}_{+4.0}$} & \multicolumn{1}{|c}{$-12.8^{-1.7}_{+2.3}$} & \multicolumn{1}{|c}{$-11.6^{-3.0}_{+4.1}$} & \multicolumn{1}{|c}{$-12.6^{-2.3}_{+4.6}$} & \multicolumn{1}{|c}{$-10.7^{-3.3}_{+4.3}$} & \multicolumn{1}{|c|}{$-11.8^{-3.1}_{+6.5}$} \\
    \hline
    \multicolumn{1}{|c}{} & \multicolumn{1}{|c}{} & \multicolumn{1}{|c}{} & \multicolumn{1}{|c}{no uncertainty} & \multicolumn{1}{|c}{} & \multicolumn{1}{|c}{-11.5} & \multicolumn{1}{|c}{-11.8} & \multicolumn{1}{|c}{-9.2} & \multicolumn{1}{|c}{-11.4} & \multicolumn{1}{|c}{-6.9} & \multicolumn{1}{|c|}{-11.0} \\ \multicolumn{1}{|c}{} & \multicolumn{1}{|c}{} & \multicolumn{1}{|c}{} & \multicolumn{1}{|c}{MBTA} & \multicolumn{1}{|c}{53} & \multicolumn{1}{|c}{$-10.3^{-4.4}$} & \multicolumn{1}{|c}{$-11.2^{-2.3}$} & \multicolumn{1}{|c}{$-9.2^{-4.1}_{+0.1}$} & \multicolumn{1}{|c}{$-10.3^{-3.9}$} & \multicolumn{1}{|c}{$-8.1^{-4.0}_{+1.3}$} & \multicolumn{1}{|c|}{$-9.0^{-5.1}$} \\ \multicolumn{1}{|c}{4.0} & \multicolumn{1}{|c}{1.4} & \multicolumn{1}{|c}{0.10} & \multicolumn{1}{|c}{equation of state} & \multicolumn{1}{|c}{40} & \multicolumn{1}{|c}{$-10.3^{-1.0}$} & \multicolumn{1}{|c}{$-11.2^{-0.4}_{+0.1}$} & \multicolumn{1}{|c}{$-9.2_{+1.2}$} & \multicolumn{1}{|c}{$-10.3^{-0.8}$} & \multicolumn{1}{|c}{$-8.1_{+2.8}$} & \multicolumn{1}{|c|}{$-8.9^{-1.8}$} \\ \multicolumn{1}{|c}{} & \multicolumn{1}{|c}{} & \multicolumn{1}{|c}{} & \multicolumn{1}{|c}{$m_{\rm ej}$, $v_{\rm ej}$} & \multicolumn{1}{|c}{} & \multicolumn{1}{|c}{$-11.6^{-1.1}_{+1.2}$} & \multicolumn{1}{|c}{$-11.9^{-0.9}_{+0.6}$} & \multicolumn{1}{|c}{$-9.7^{-1.6}_{+2.0}$} & \multicolumn{1}{|c}{$-11.2^{-0.7}_{+0.3}$} & \multicolumn{1}{|c}{$-7.9^{-2.3}_{+3.1}$} & \multicolumn{1}{|c|}{$-10.7^{-0.2}_{+0.2}$} \\ \multicolumn{1}{|c}{} & \multicolumn{1}{|c}{} & \multicolumn{1}{|c}{} & \multicolumn{1}{|c}{$X_{\rm lan}$} & \multicolumn{1}{|c}{} & \multicolumn{1}{|c}{$-12.2^{-0.5}_{+4.8}$} & \multicolumn{1}{|c}{$-11.0^{-2.7}_{+0.6}$} & \multicolumn{1}{|c}{$-11.0^{-1.1}_{+4.9}$} & \multicolumn{1}{|c}{$-9.8^{-3.7}_{+2.8}$} & \multicolumn{1}{|c}{$-9.8^{-1.8}_{+5.9}$} & \multicolumn{1}{|c|}{$-9.1^{-3.7}_{+5.9}$} \\ \multicolumn{1}{|c}{} & \multicolumn{1}{|c}{} & \multicolumn{1}{|c}{} & \multicolumn{1}{|c}{all} & \multicolumn{1}{|c}{27} & \multicolumn{1}{|c}{$-10.4^{-2.5}_{+0.1}$} & \multicolumn{1}{|c}{$-11.2^{-1.7}$} & \multicolumn{1}{|c}{$-9.3^{-2.5}_{+0.1}$} & \multicolumn{1}{|c}{$-10.3^{-2.5}_{+0.1}$} & \multicolumn{1}{|c}{$-8.2^{-2.8}_{+0.1}$} & \multicolumn{1}{|c|}{$-8.9^{-3.0}_{+0.1}$} \\
    \hline
    \multicolumn{1}{|c}{} & \multicolumn{1}{|c}{} & \multicolumn{1}{|c}{} & \multicolumn{1}{|c}{no uncertainty} & \multicolumn{1}{|c}{} & \multicolumn{1}{|c}{-15.6} & \multicolumn{1}{|c}{-13.7} & \multicolumn{1}{|c}{-14.4} & \multicolumn{1}{|c}{-14.6} & \multicolumn{1}{|c}{-13.4} & \multicolumn{1}{|c|}{-15.2} \\ \multicolumn{1}{|c}{} & \multicolumn{1}{|c}{} & \multicolumn{1}{|c}{} & \multicolumn{1}{|c}{MBTA} & \multicolumn{1}{|c}{54} & \multicolumn{1}{|c}{$-10.7^{-5.2}_{+0.3}$} & \multicolumn{1}{|c}{$-11.2^{-3.0}$} & \multicolumn{1}{|c}{$-9.2^{-5.5}$} & \multicolumn{1}{|c}{$-10.5^{-4.3}_{+0.2}$} & \multicolumn{1}{|c}{$-8.1^{-5.8}$} & \multicolumn{1}{|c|}{$-10.1^{-4.9}_{+1.2}$} \\ \multicolumn{1}{|c}{4.0} & \multicolumn{1}{|c}{1.4} & \multicolumn{1}{|c}{0.70} & \multicolumn{1}{|c}{equation of state} & \multicolumn{1}{|c}{100} & \multicolumn{1}{|c}{$-14.9^{-0.7}_{+1.2}$} & \multicolumn{1}{|c}{$-13.5^{-0.3}_{+0.6}$} & \multicolumn{1}{|c}{$-13.5^{-0.8}_{+1.4}$} & \multicolumn{1}{|c}{$-14.2^{-0.5}_{+0.7}$} & \multicolumn{1}{|c}{$-12.3^{-1.1}_{+1.7}$} & \multicolumn{1}{|c|}{$-14.3^{-0.9}_{+1.1}$} \\ \multicolumn{1}{|c}{} & \multicolumn{1}{|c}{} & \multicolumn{1}{|c}{} & \multicolumn{1}{|c}{$m_{\rm ej}$, $v_{\rm ej}$} & \multicolumn{1}{|c}{} & \multicolumn{1}{|c}{$-15.1^{-0.5}_{+0.7}$} & \multicolumn{1}{|c}{$-13.8^{-0.6}_{+1.0}$} & \multicolumn{1}{|c}{$-13.9^{-0.5}_{+1.2}$} & \multicolumn{1}{|c}{$-14.6^{-0.3}_{+1.0}$} & \multicolumn{1}{|c}{$-13.3^{-0.4}_{+1.9}$} & \multicolumn{1}{|c|}{$-14.3^{-1.1}_{+0.3}$} \\ \multicolumn{1}{|c}{} & \multicolumn{1}{|c}{} & \multicolumn{1}{|c}{} & \multicolumn{1}{|c}{$X_{\rm lan}$} & \multicolumn{1}{|c}{} & \multicolumn{1}{|c}{$-16.1^{-0.1}_{+5.3}$} & \multicolumn{1}{|c}{$-13.3^{-1.9}_{+0.1}$} & \multicolumn{1}{|c}{$-15.0^{-0.4}_{+5.0}$} & \multicolumn{1}{|c}{$-14.2^{-1.7}_{+0.2}$} & \multicolumn{1}{|c}{$-14.4^{-0.6}_{+5.0}$} & \multicolumn{1}{|c|}{$-14.3^{-1.9}_{+0.6}$} \\ \multicolumn{1}{|c}{} & \multicolumn{1}{|c}{} & \multicolumn{1}{|c}{} & \multicolumn{1}{|c}{all} & \multicolumn{1}{|c}{44} & \multicolumn{1}{|c}{$-10.4^{-5.2}_{+0.2}$} & \multicolumn{1}{|c}{$-11.2^{-3.2}_{+0.1}$} & \multicolumn{1}{|c}{$-9.3^{-5.5}_{+0.2}$} & \multicolumn{1}{|c}{$-10.3^{-4.5}_{+0.1}$} & \multicolumn{1}{|c}{$-8.3^{-6.0}_{+0.2}$} & \multicolumn{1}{|c|}{$-9.0^{-5.8}_{+0.2}$} \\
    \hline
    \multicolumn{1}{|c}{} & \multicolumn{1}{|c}{} & \multicolumn{1}{|c}{} & \multicolumn{1}{|c}{no uncertainty} & \multicolumn{1}{|c}{} & \multicolumn{1}{|c}{-11.7} & \multicolumn{1}{|c}{-11.9} & \multicolumn{1}{|c}{-9.4} & \multicolumn{1}{|c}{-11.7} & \multicolumn{1}{|c}{-7.1} & \multicolumn{1}{|c|}{-11.3} \\ \multicolumn{1}{|c}{} & \multicolumn{1}{|c}{} & \multicolumn{1}{|c}{} & \multicolumn{1}{|c}{MBTA} & \multicolumn{1}{|c}{16} & \multicolumn{1}{|c}{$-10.3^{-3.2}$} & \multicolumn{1}{|c}{$-11.2^{-1.5}$} & \multicolumn{1}{|c}{$-9.2^{-2.7}$} & \multicolumn{1}{|c}{$-10.3^{-2.8}$} & \multicolumn{1}{|c}{$-8.1^{-2.2}$} & \multicolumn{1}{|c|}{$-9.0^{-3.7}$} \\ \multicolumn{1}{|c}{4.0} & \multicolumn{1}{|c}{2.0} & \multicolumn{1}{|c}{0.70} & \multicolumn{1}{|c}{equation of state} & \multicolumn{1}{|c}{46} & \multicolumn{1}{|c}{$-10.3^{-2.7}$} & \multicolumn{1}{|c}{$-11.2^{-1.2}$} & \multicolumn{1}{|c}{$-9.2^{-2.1}_{+0.4}$} & \multicolumn{1}{|c}{$-10.3^{-2.5}$} & \multicolumn{1}{|c}{$-8.1^{-1.5}_{+1.8}$} & \multicolumn{1}{|c|}{$-9.0^{-3.3}$} \\ \multicolumn{1}{|c}{} & \multicolumn{1}{|c}{} & \multicolumn{1}{|c}{} & \multicolumn{1}{|c}{$m_{\rm ej}$, $v_{\rm ej}$} & \multicolumn{1}{|c}{} & \multicolumn{1}{|c}{$-11.8^{-1.2}_{+1.1}$} & \multicolumn{1}{|c}{$-12.0^{-0.9}_{+0.8}$} & \multicolumn{1}{|c}{$-9.9^{-1.7}_{+1.9}$} & \multicolumn{1}{|c}{$-11.4^{-0.8}_{+0.3}$} & \multicolumn{1}{|c}{$-8.0^{-2.4}_{+3.0}$} & \multicolumn{1}{|c|}{$-10.9^{-0.3}_{+0.2}$} \\ \multicolumn{1}{|c}{} & \multicolumn{1}{|c}{} & \multicolumn{1}{|c}{} & \multicolumn{1}{|c}{$X_{\rm lan}$} & \multicolumn{1}{|c}{} & \multicolumn{1}{|c}{$-12.3^{-0.5}_{+4.9}$} & \multicolumn{1}{|c}{$-11.3^{-2.4}_{+0.6}$} & \multicolumn{1}{|c}{$-11.1^{-1.1}_{+5.1}$} & \multicolumn{1}{|c}{$-10.2^{-3.5}_{+2.2}$} & \multicolumn{1}{|c}{$-10.0^{-1.7}_{+6.0}$} & \multicolumn{1}{|c|}{$-9.4^{-3.7}_{+4.8}$} \\ \multicolumn{1}{|c}{} & \multicolumn{1}{|c}{} & \multicolumn{1}{|c}{} & \multicolumn{1}{|c}{all} & \multicolumn{1}{|c}{14} & \multicolumn{1}{|c}{$-10.4^{-1.7}_{+0.1}$} & \multicolumn{1}{|c}{$-11.2^{-1.9}$} & \multicolumn{1}{|c}{$-9.3^{-1.7}_{+0.1}$} & \multicolumn{1}{|c}{$-10.3^{-2.9}_{+0.1}$} & \multicolumn{1}{|c}{$-8.2^{-2.0}_{+0.1}$} & \multicolumn{1}{|c|}{$-8.9^{-3.5}_{+0.1}$} \\
    \hline
    \end{tabular}
    \caption{\textit{HasEjecta} and absolute magnitude, assigned with upper (10th percentile) and lower (90th percentile) limits, for different binaries and different  sources of uncertainty. The real parameters, $m_1^{\rm fixed}$, $m_2^{\rm fixed}$, $\chi_{\rm eff}^{\rm fixed}$, of the binaries are provided in the first three columns. Always when there is no marginalization over the entire set of $2396$ equations of state, ${\rm EOS}^{\rm fixed}$ corresponds to $\{\gamma\} = (1.4777, -0.3225, 0.0694, -0.0046)$ and associates to a $1.4M_\odot$ neutron star, a radius of $13.0$ km and a tidal deformability of 663, while the maximum Tolman-Oppenheimer-Volkov mass is $ 2.43M_\odot$.
    The fourth column indicates either the unique source of uncertainty  (MBTA; equation of state; $m_{\rm ej}, v_{\rm ej}$; $X_{\rm lan}$), if no uncertainty was considered, or if all the combined uncertainty sources have been taken into account. The absolute magnitudes, and their error bars are reported for the $\mathit{\mathbf{g}}$ and $\mathit{\mathbf{K}}$ photometric filters, after the first, second and third day, with respect to the coalescence time. All the simulations have been realized with \textit{Model I}.}
    \label{tab:uncertainty_sources}
\end{table*}

Table~\ref{tab:uncertainty_sources} summarizes these results for five binaries. The parameters have been chosen in such a way that, based on our knowledge to date, the binaries are: a BNS ($m_1^{\rm fixed} = 1.6 M_{\odot} $, $m_2^{\rm fixed} = 1.4 M_{\odot}$ , $\chi_{\rm eff}^{\rm fixed} = 0.01$); a system which, depending on the neutron star equation of state, is either a BNS or a NSBH ($m_1^{\rm fixed} = 2.0 M_{\odot} $, $m_2^{\rm fixed} = 1.4 M_{\odot}$ , $\chi_{\rm eff}^{\rm fixed} = 0.01$); a low spin NSBH ($m_1^{\rm fixed} = 4.0 M_{\odot} $, $m_2^{\rm fixed} = 1.4 M_{\odot}$ , $\chi_{\rm eff}^{\rm fixed} = 0.1$); a high spin NSBH ($m_1^{\rm fixed} = 4.0 M_{\odot} $, $m_2^{\rm fixed} = 1.4 M_{\odot}$ , $\chi_{\rm eff}^{\rm fixed} = 0.7$); a system which, depending on the supra-nuclear matter equation of state, is either a NSBH or a BBH, and has high spin ($m_1^{\rm fixed} = 4.0 M_{\odot} $, $m_2^{\rm fixed} = 2.0 M_{\odot}$ , $\chi_{\rm eff}^{\rm fixed} = 0.7$).
First of all, we remark the expected behavior of lightcurve uncertainties increasing with time. The most important source of uncertainty turns to be our ignorance about the chemical composition of the ejecta. Letting $X_{\rm lan}$ vary within $[10^{-9}, 10^{-1}]$ is responsible of a difference of up to five magnitudes at the end of the first day. Then the second main source of uncertainty seems to be the inaccurate GW strain  measurement by the low latency-pipelines. The errors are greater when the system is not undoubtedly a BNS. There are at least two simple explanations for this feature: the high uncertainty on the mass ratio implies considering systems of different types (BNS and NSBH; NSBH and BBH); in the case of high effective spin $\chi_{\rm eff}^{\rm fixed}$, our choice of the variation interval $[-\chi_{\rm eff}^{\rm fixed}, \min{(3\chi_{\rm eff}^{\rm fixed}, 1)}]$ has a non-negligible impact on the mass ejecta, as we show in Figure~\ref{fig:nsbh_ejected}. The last two sources of uncertainty are the equation of state marginalization and the ejecta fit errors. As expected, the effects of the equation of state marginalization are more substantial when one of the binary components has a mass of about $2M_\odot$. In such a case, by varying the equation of state, we change the type of the compact object. Therefore, the low-latency measurement errors have a big influence on \textit{HasEjecta}, especially when the system is high spinning and has a non-negligible probability to be a NSBH. Similarly, the effect of the equation of state marginalization on \textit{HasEjecta} is important, as was previously highlighted in Figure~\ref{fig:bns_ejected}. It is worth noting that when all the error sources are considered, the corresponding uncertainty is not the simple sum of the independent errors, since we allow for compensation effect.

\begin{figure}
 \includegraphics[width=0.5\textwidth]{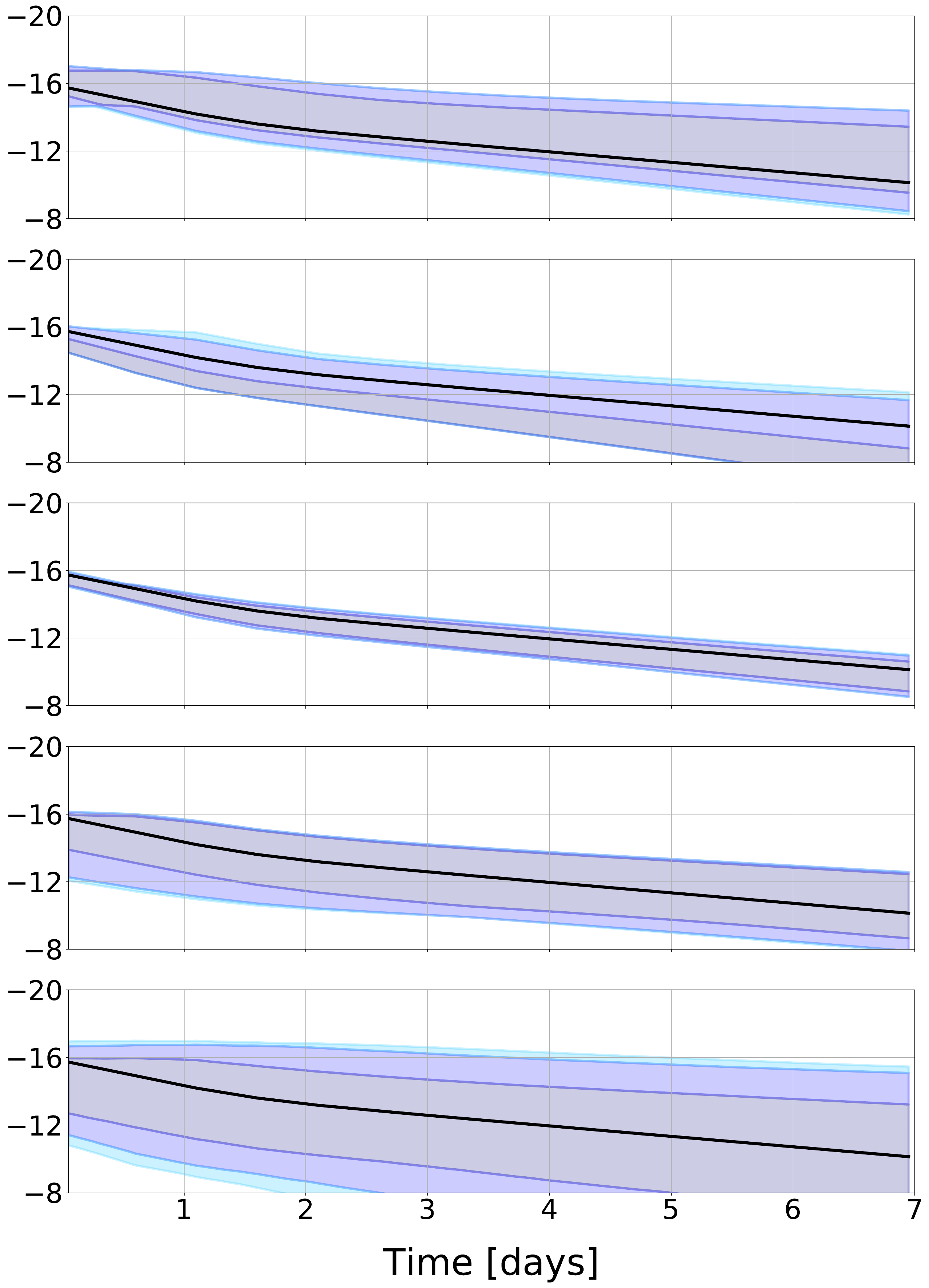}
  \caption{Absolute magnitude versus time for a BNS system with $(m_1^{\rm fixed}, m_2^{\rm fixed}, \chi_{\rm eff}^{\rm fixed}) = (1.6M_\odot, 1.4M_\odot, 0.01)$. The blue filled bands represent the predicted lightcurves assigned with errors. The sources of uncertainty are, from the top panel to the bottom panel, the MBTA low-latency pipeline, the equation of state marginalization, the mass and velocity of the ejecta, the lanthanide fraction, and all the uncertainty sources combined. In all panels, the black curve represents the same predicted lightcurve when there is no uncertainty at all. In this case, the equation of state is ${\rm EOS}^{\rm fixed}$, parameterized by $\{\gamma\} = (1.4777, -0.3225, 0.0694, -0.0046)$,  and the lanthanide fraction is fixed to $X_{\rm lan}^{\rm fixed} = 10^{-4}$. In the contour plots, $1\sigma$, $2\sigma$ and $3\sigma$ are indicated in shades of blue, from darkest to lightest. For these simulations we used \textit{Model I}.}
 \label{fig:uncertainty_on_BNS}
\end{figure}

In Figure~\ref{fig:uncertainty_on_BNS}, we show an example of a BNS lightcurve contour, illustrating the time evolution of the absolute magnitude in the  $\mathit{\mathbf{r}}$ photometric band, corresponding to the different sources of uncertainty. Except the case of only equation of state marginalization uncertainty, in all the other cases the true value of the absolute magnitude is included inside the $1\sigma$ error. Figure~\ref{fig:uncertainty_on_BNS} points out again that the main limitations of our pipeline are: the lack of knowledge regarding the ejecta composition and the imprecise constraint of the binary components parameters. In contrary, the $\mathit{\mathbf{r}}$ filter $3\sigma$ error bar, corresponding to the eventual imprecise ejecta fits, spreads over less than 2 magnitudes after 3 days succeeding the compact object coalescence.

\section{Demonstration on real examples}
\label{sec:O2/O3 events}
In this section, we demonstrate the output of our tool on the following O2/O3a LIGO-Virgo GW events: GW170817~\citep{PhysRevLett.119.161101}, GW190425~\citep{Abbott:2020uma}, and GW190814~\citep{Abbott_2020}. 
GW170817 and GW190425 are BNSs, while GW190814 is either a NSBH or a BBH; cf.~the discussion in e.g.~\citep{Abbott_2020,Essick:2020ghc,Most:2020bba,Tews:2020ylw,Tan:2020ics}. It is worth mentioning that more than 98\% of the time needed for the code to run is used in the equation of state marginalization (presented in Section~\ref{Dynamical Ejecta}) and lightcurve generation (presented in Section~\ref{sec:surrogates}) processes. More precisely with a single \textit{E5-2698 v4} processor, we need on average 6.204s for the equation of state marginalization and 0.198s (respectively 0.471s) for the computation of a \textit{Model I} (respectively \textit{Model II}) lightcurve, if only one core is used. In this case the total necessary time to convert the input low-latency data into kilonova lightcurves is around 198s + $6.204s \times n_{\rm templates}$ (respectively 471s + $6.204s \times n_{\rm templates}$) when \textit{Model I} (respectively \textit{Model II}) is used. In the preceding expression $n_{\rm templates}$ is the number of input MBTA templates, which typically is $O(10)$. We note, though, that this computation is easily parallelizable and latency could be reduced. For example, when the same processor is used with 8 cores, the times required for the equation of state marginalization, \textit{Model I} and \textit{Model II} lightcurve computations become 0.975s, 0.059s and 0.249s which means that the overall code is executed in around 59s + 0.975s $\times n_{\rm templates}$ (respectively 249s + 0.975s $\times n_{\rm templates}$) if \textit{Model I} (respectively \textit{Model II}) is employed. 


\subsection{Comparison of ejecta mass and \textit{HasEjecta}}
\begin{figure}
    \centering
    \includegraphics[width=0.45\textwidth]{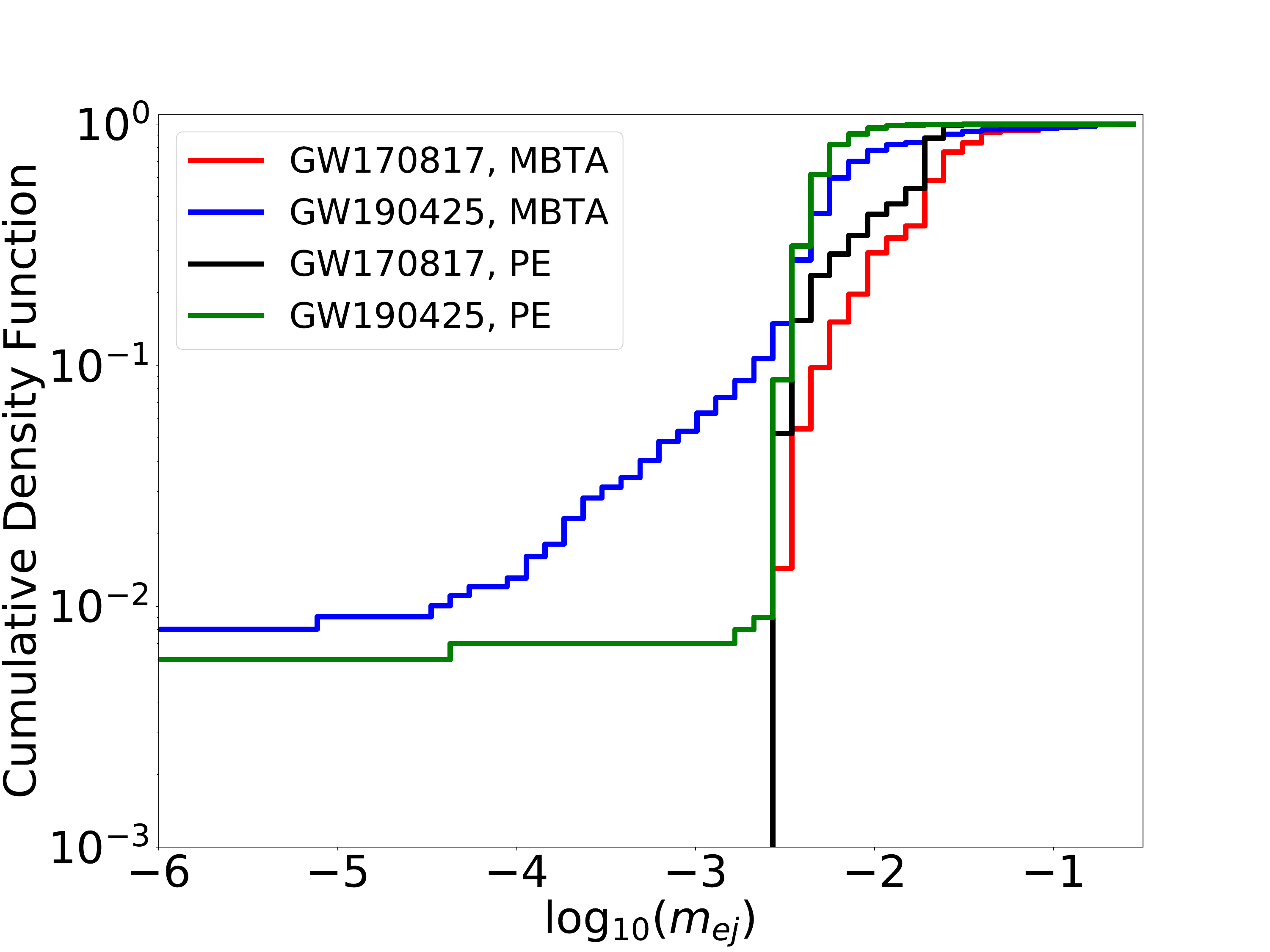}
    \caption{Cumulative distribution function of the ejected mass $m_{\rm ej}$ for the following GW triggers: GW170817, GW190425. There are both the low-latency results (the input data is represented by the MBTA weighted templates) and the PE results (the input data is represented by the offline PE~\citep{Veitch:2014wba} posteriors).
    }
    \label{fig:mej_nonzero}
\end{figure}

The definition of \textit{HasEjecta} is similar to that of \textit{HasRemnant}~\citep{Chatterjee:2019avs}, a low-latency data based product provided by the LVC. In Table~\ref{tab:has_ejecta}, we compare \textit{HasRemnant} and \textit{HasEjecta} calculated in two ways: (i) the MBTA samples described in Section~\ref{sec:MBTA_uncertainties}, and (ii) sampling the waveform model posteriors of PE results. From this table, the three quantities give consistent results.
From the list of three events mentioned at the beginning of this section, only two of them (GW170817 and GW190425) have non-negligible \textit{HasEjecta}.
\begin{table}
\caption[]{
Values of \textit{HasRemnant} (second column) and \textit{HasEjecta}, based on both MBTA (third column) and PE  (fourth column) samples, for the O2/O3a compact binary coalescence events: GW170817, GW190425 and GW190814. The values of \textit{HasRemnant} for GW190425 and GW190814 triggers are taken from \citep{2019GCN.24168....1L} and \citep{2019GCN.25324....1L}. Given the fact that \textit{HasRemnant} has been introduced at the beginning of 03, for the GW170817 \textit{HasRemnant} we had to assume the same value as for \textit{EM-Bright} reported in~\citep{AbEA2019}. Regarding the PE results, the samples used for GW170817 (respectively GW190425 and GW190814) are the ones introduced in~\cite{LIGOScientific:2018mvr} (respectively in~\cite{Abbott:2020niy}).

}
\label{tab:has_ejecta}
\begin{tabular}{lccc}
Event & HasRemnant & MBTA HasEjecta & PE HasEjecta\\
\hline
 GW170817 &  100\% & 100\% & 100\%\\ 
 GW190425 & > 99\% & 98\% & >99\%\\
 GW190814 & < 1\% & 0\% & 0\%\\ 
\hline
\end{tabular}
\end{table} 
Therefore in Figure~\ref{fig:mej_nonzero} there is an illustration of mass ejecta $m_\mathrm{ej}$ distribution for GW170817 and GW190425. This figure suggests that the low-latency based method presented in this paper reproduces fairly well the predictions one could get by means of the offline PE posteriors, however, further studies of other events are required for a final conclusion.      

\subsection{Lightcurve predictions}
\label{sec:apparent_magnitude}

\begin{figure*}
 \includegraphics[angle=90, width=\textwidth, height=0.4\textheight]{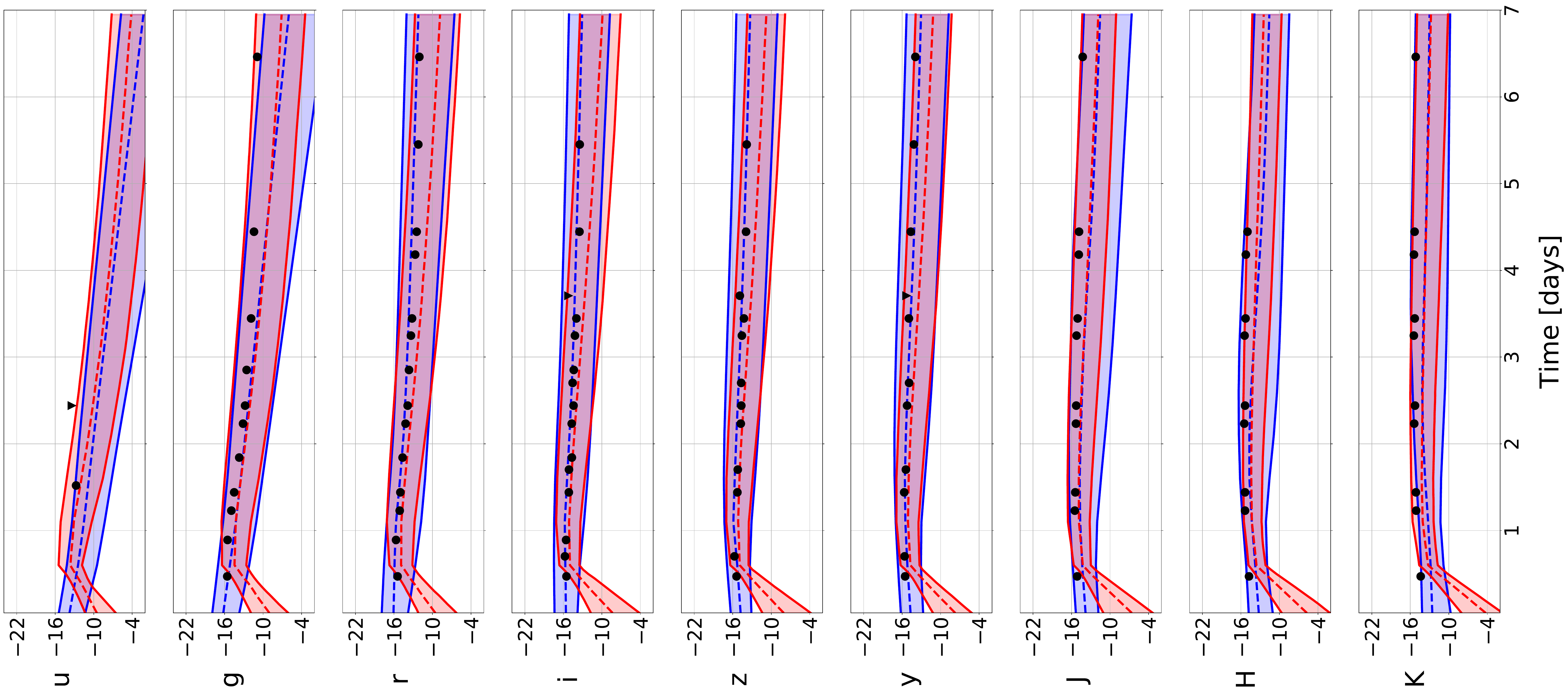} 
\includegraphics[angle=90,  width=\textwidth, height=0.4\textheight]{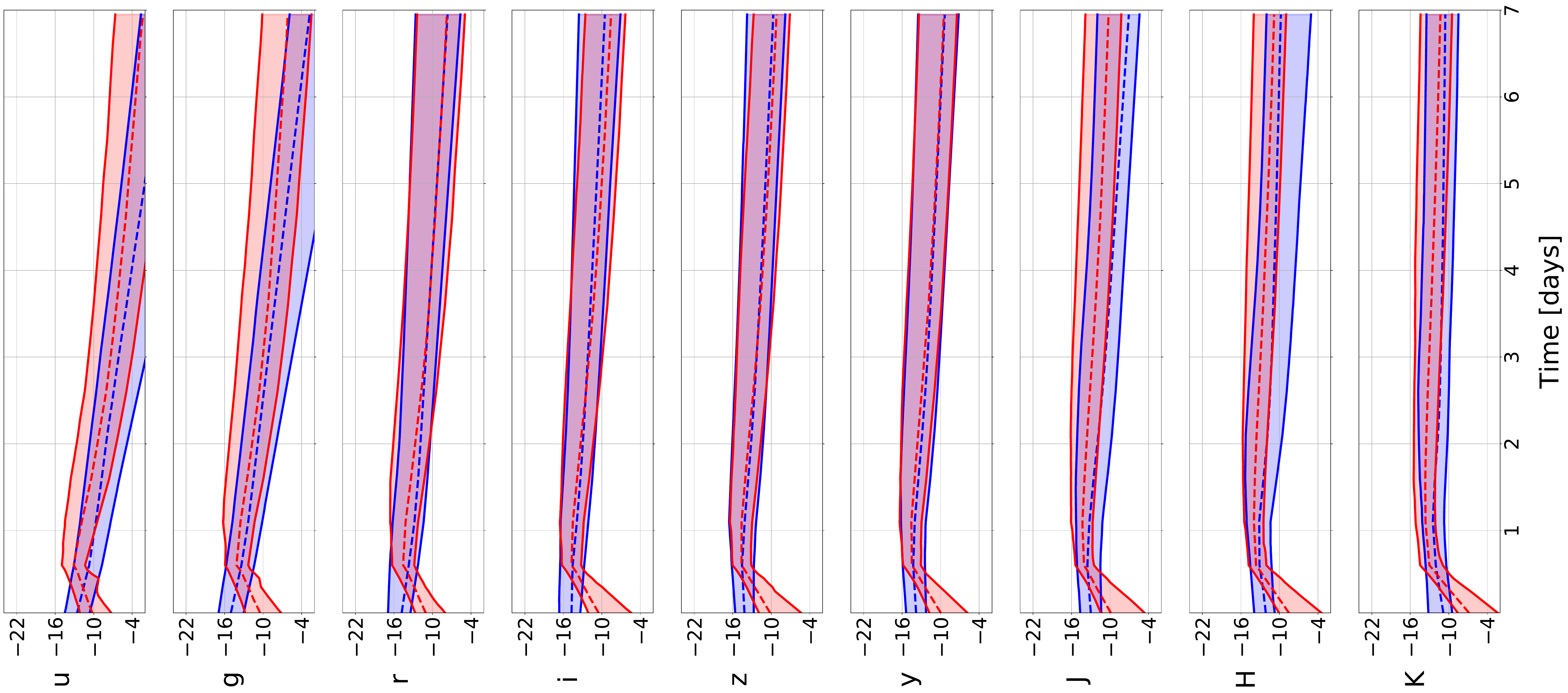}
  \caption{Kilonova lightcurves (absolute magnitude versus time) predicted by \textit{Model I} (in blue) and \textit{Model II} (in red) for $u, g, r, i, z, y, J, H, K$ photometric bands. The predicted lightcurve for GW170817 (top) and for GW190425 (bottom). For each filter there are three curves plotted: the upper solid line, the middle dashed line and respectively the lower solid line represent the 90th, 50th and respectively 10th percentiles. In the case of GW170817 the circle and triangle symbols in black illustrate the measured optical points of AT2017gfo. The circle symbols are measure points with finite uncertainty, while triangle symbols are upper limits. The prior $X_\mathrm{lan} = 10
 ^{-4}$ 
 (respectively $\Phi = 45^\circ$ and $\cos{\theta_\mathrm{inc}}$ uniform in [0,1]) is used for \textit{Model I} (respectively \textit{Model II}).
 }
 \label{fig:Kasen_Bulla_lightcurves}
\end{figure*}

In Figure~\ref{fig:Kasen_Bulla_lightcurves}, we illustrate the corresponding lightcurves employing both \textit{Model I} and \textit{Model II}. Regarding \textit{Model I}, we set the lanthanide fraction for this analysis to $X_\mathrm{lan} = 10^{-4}$ (consistent with the results presented in~\cite{Coughlin:2018miv}), 
while concerning \textit{Model II}, we use a uniform prior in $\cos\theta_\mathrm{inc}$, for the inclination angle $\theta_\mathrm{inc}$, and we fix $\Phi = 45^\circ$ (consistent with the results presented in~\citep{Dietrich:2020lps}). A significant difference between the two models, clearly highlighted by Figure~\ref{fig:Kasen_Bulla_lightcurves}, concerns the first half day following the kilonova. \textit{Model I} lightcurves present a small (negative for the lower wavelengths and positive for the higher wavelengths) slope, while \textit{Model II} has a more pronounced raising shape. 

One can observe a good agreement between the real data and our predictions in the case of GW170817. Almost all the observational points are included in between the upper and lower limits in the case of \textit{Model I}, whereas the predictions from \textit{Model II} are missing a few more points in the first half day. Nevertheless the agreement could be strengthened by choosing other parameters ($X_\mathrm{lan}$, $\theta_\mathrm{inc}$, $\Phi$), which are not constrained by the GW data and/or our compact objects understanding. Also, this suggests that the uncertainty presented here could be significantly underestimated because we fixed these parameters. The \textit{Model I} predictions for GW190425 show that it is less bright than GW170817. Indeed, predictions for GW190425 are at least 1 magnitude higher (i.e., dimmer) in almost all photometric bands after only 1 hour, and at least 3 magnitude higher after 3 days. This fact corroborated with the broad skymap localization~\citep{2019GCN.24168....1L} could explain the non-detection of an EM counterpart for GW190425. 

\begin{figure}
 \includegraphics[width=0.5\textwidth]{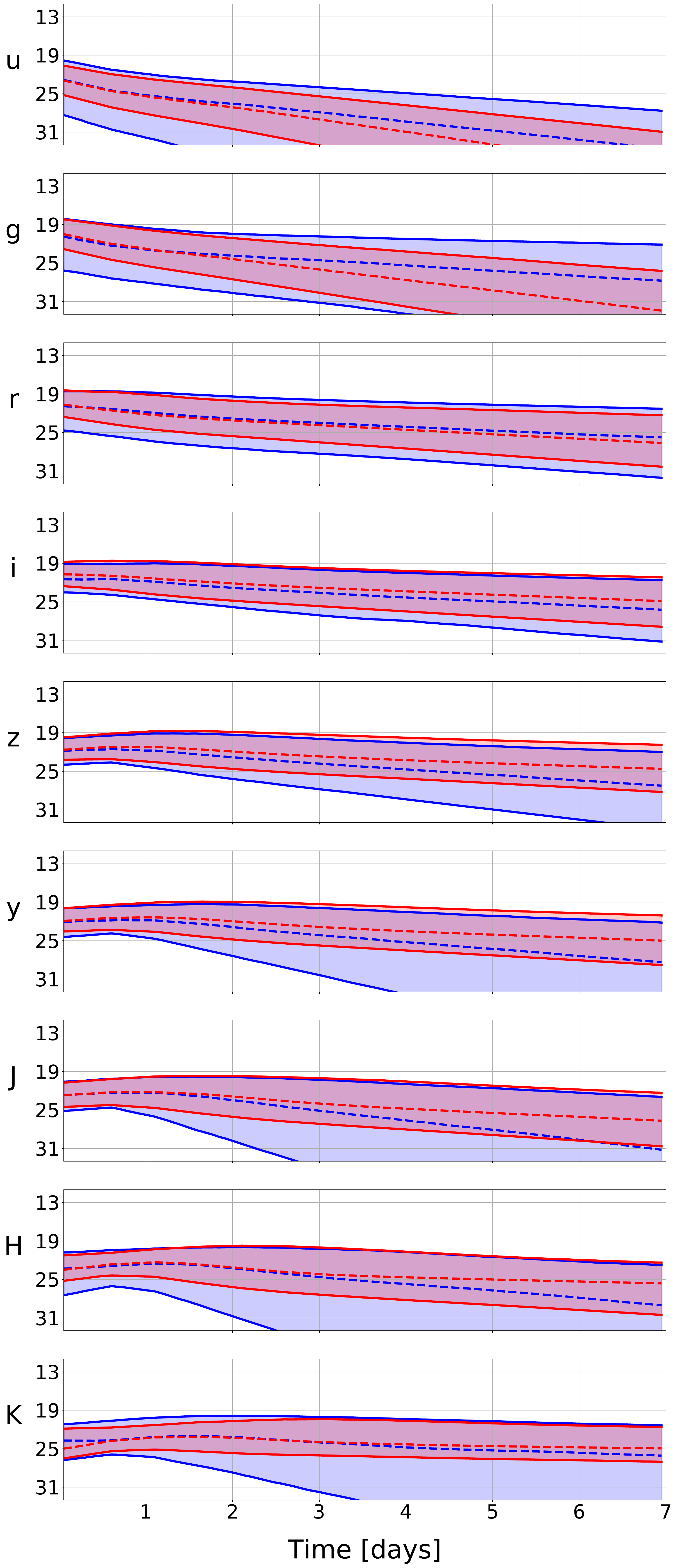}
  \caption{Apparent magnitude versus time based on the offline PE posteriors and LALInference~\citep{Veitch:2014wba} skymap. The presented event is GW190425 and the lightcurves in blue (respectively red) are predicted by \textit{Model I} with prior $\log_{10}X_{\rm lan}$ uniform in $[-9, -1]$ (respectively $X_{\rm lan} = 10^{-4}$). }
 \label{fig:app_bayestar_MBTA_Kasen_190425}
\end{figure}

In order to convert absolute magnitude in apparent magnitude, we use the distance in the form of the distance modulus: $m - M = 5\log_{10}d_L - 5$, where $m, M, d_L$ are the apparent magnitude, the absolute magnitude, and the luminosity distance expressed in units of pc. Here, we will use LIGO-Virgo-Kagra based low-latency products which contain the required distance information. One of the data released in low-latency is the Bayestar skymap~\citep{SiPr2016}; this skymap provides an array of sky coordinates, each of them being assigned with a localization probability, a luminosity distance and a distance uncertainty. From this skymap a mean distance is calculated.
It is worth mentioning that the luminosity distance uncertainty is not taken into account for the calculation of the lightcurve contours. For example the Bayestar distance relative error for GW170817, GW190425, and GW190814 is less than 0.3 which translates to an uncertainty on the apparent magnitude of less than 0.7 magnitudes. Nonetheless, this value is not negligible, as it represents  $\sim20\%$ of the total error budget. This suggests that the uncertainty stated here is underestimated. 
In  Figure~\ref{fig:app_bayestar_MBTA_Kasen_190425} there is an example of contours representing the evolution of the apparent magnitude with time. This figure shows again that the lack of information concerning the chemical composition of the ejecta is responsible of a large uncertainty. Also this figure emphasizes that whatever the lanthanide fraction $X_{\rm lan}\in [10^{-9}, 10^{-1}]$, the output for GW190425 predicts a kilonova with apparent magnitude higher than 21.2 (respectively 20.2) in $g$ (respectively $r$) photometric band after only one day, which is in agreement with the ZTF observational data based upper limits presented in~\cite{Coughlin:2019xfb}.


\section{Conclusion}
\label{sec:conclusion}

In this paper, we present a tool aimed to predict kilonova lightcurves based on the low-latency data products.
We propose a way to take advantage of the multiple templates around the preferred event found by the low-latency pipelines and show how to predict mass ejecta from preliminary estimated of the binary parameters. We demonstrate the procedure on GW candidates, computing the ejecta probability (\textit{HasEjecta}) reported by our tool and comparing to the value of  \textit{HasRemnant} currently released by the LIGO-Virgo-Kagra Collaborations. We then propose two ways to convert mass ejecta and other parameters such as ejecta velocity, lanthanide fraction, binary inclination angle and half opening angle of the lanthanide-rich ejecta component into kilonova lightcurves. The different sources of errors are evaluated. It turns out that the knowledge uncertainty we have about the chemical composition of the ejecta is the principal limitation of the method, while the mass ejecta fits errors have the smallest impact on the lightcurve output. 
We compare our predicted lightcurve with the only kilonova counterpart observed to date, i.e., AT2017gfo, showing consistency with those results. Finally, we suggest how to convert absolute magnitude to apparent magnitude by means of Bayestar skymap. This method can be used during the next observing run O4 as a utility to better inform the EM community concerning the characteristics of the kilonova signal they are trying to catch.

Improvements to the tool can be envisaged. A better treatment of the input low-latency data of LIGO-Virgo-Kagra could be considered, in particular, the availability of low-latency parameter estimation results might be of importance, since the uncertainties on the individual mass are non-negligible and only the chirp mass is quite well measured. On the other hand, the mass ratio for the existing binary population in the Universe considerably improved during the last years.
As a consequence, one way to reduce errors could be to consider only the chirp mass from templates~\citep{Margalit:2019dpi}, but use the mass ratio based on the observed binary population~\citep[e.g.,][]{PhysRevD.81.084029, Abbott:2020gyp,Essick:2020ghc,Fishbach:2020ryj,Fishbach_2020}. 
Equally, one could provide lightcurve estimates conditioned on the direction to the source, which will probably be what an EM observer would want. Such a development should be easily implementable given the format of the actual Bayestar and LALInference skymaps.
Likewise, more counterparts to binary compact mergers in the next years will improve our understanding of the equation of state of supra-nuclear-dense matter and potentially of the ejecta composition and geometry of the different ejecta components. Thereafter, priors like the lanthanide fraction and/or the half-opening angle of some ejecta component, needed for the computation of lightcurve made by the surrogates, could be addressed more accurately. \\

\emph{Acknowledgements.} 
M.C. acknowledges support from the National Science Foundation with grant number PHY-2010970. N.C.  acknowledges  support  from  the National Science Foundation with grant number PHY-1806990. 
M.B. acknowledges support from the Swedish Research Council (Reg. no. 2020-03330). S.A is supported by the CNES Postdoctoral Fellowship at Laboratoire AstroParticule et Cosmologie.
R.E. was supported by the Kavli Institute for Cosmological Physics and the Perimeter Institute for Theoretical Physics.
The Kavli Institute for Cosmological Physics at the University of Chicago is supported through an endowment from the Kavli Foundation and its founder Fred Kavli.
Research at Perimeter Institute is supported in part by the Government of Canada through the Department of Innovation, Science and Economic Development Canada and by the Province of Ontario through the Ministry of Colleges and Universities.
P.L. is supported by National Science Foundation award PHY-1836734 and by a gift from the Dan Black Family Foundation to the Gravitational-Wave Physics \& Astronomy Center.

We thank our colleagues from the MBTA team for sharing this pipeline and for useful discussions.



\section*{Data Availability}
The data underlying this article are derived from public code found here: \href{https://github.com/mcoughlin/gwemlightcurves}{https://github.com/mcoughlin/gwemlightcurves}. The simulations resulting will be shared on reasonable request to the corresponding author.

\bibliographystyle{mnras}
\bibliography{refs.bib} 

\begin{thebibliography}{}
\makeatletter
\relax
\def\mn@urlcharsother{\let\do\@makeother \do\$\do\&\do\#\do\^\do\_\do\%\do\~}
\def\mn@doi{\begingroup\mn@urlcharsother \@ifnextchar [ {\mn@doi@}
  {\mn@doi@[]}}
\def\mn@doi@[#1]#2{\def\@tempa{#1}\ifx\@tempa\@empty \href
  {http://dx.doi.org/#2} {doi:#2}\else \href {http://dx.doi.org/#2} {#1}\fi
  \endgroup}
\def\mn@eprint#1#2{\mn@eprint@#1:#2::\@nil}
\def\mn@eprint@arXiv#1{\href {http://arxiv.org/abs/#1} {{\tt arXiv:#1}}}
\def\mn@eprint@dblp#1{\href {http://dblp.uni-trier.de/rec/bibtex/#1.xml}
  {dblp:#1}}
\def\mn@eprint@#1:#2:#3:#4\@nil{\def\@tempa {#1}\def\@tempb {#2}\def\@tempc
  {#3}\ifx \@tempc \@empty \let \@tempc \@tempb \let \@tempb \@tempa \fi \ifx
  \@tempb \@empty \def\@tempb {arXiv}\fi \@ifundefined
  {mn@eprint@\@tempb}{\@tempb:\@tempc}{\expandafter \expandafter \csname
  mn@eprint@\@tempb\endcsname \expandafter{\@tempc}}}

\bibitem[\protect\citeauthoryear{{Abbott et al.}}{{Abbott et
  al.}}{2017a}]{AbEA2017b}
{Abbott et al.} 2017a, \mn@doi [Phys. Rev. Lett.]
  {10.1103/PhysRevLett.119.161101}, 119, 161101

\bibitem[\protect\citeauthoryear{Abbott et~al.,}{Abbott
  et~al.}{2017b}]{PhysRevLett.119.161101}
Abbott B.~P.,  et~al., 2017b, \mn@doi [Phys. Rev. Lett.]
  {10.1103/PhysRevLett.119.161101}, 119, 161101

\bibitem[\protect\citeauthoryear{{Abbott et al.}}{{Abbott et
  al.}}{2017c}]{AbEA2017f}
{Abbott et al.} 2017c, The Astrophysical Journal Letters, 848, L12

\bibitem[\protect\citeauthoryear{Abbott et~al.,}{Abbott
  et~al.}{2017d}]{Abbott_2017}
Abbott B.~P.,  et~al., 2017d, \mn@doi [The Astrophysical Journal]
  {10.3847/2041-8213/aa920c}, 848, L13

\bibitem[\protect\citeauthoryear{Abbott et~al.}{Abbott
  et~al.}{2018a}]{Abbott:2018exr}
Abbott B.~P.,  et~al., 2018a, {arXiv: 1805.11581}

\bibitem[\protect\citeauthoryear{Abbott et~al.,}{Abbott
  et~al.}{2018b}]{article}
Abbott B.,  et~al., 2018b, \mn@doi [Living Reviews in Relativity]
  {10.1007/s41114-018-0012-9}, 21

\bibitem[\protect\citeauthoryear{Abbott et~al.}{Abbott
  et~al.}{2019}]{LIGOScientific:2018mvr}
Abbott B.~P.,  et~al., 2019, \mn@doi [Phys. Rev. X]
  {10.1103/PhysRevX.9.031040}, 9, 031040

\bibitem[\protect\citeauthoryear{Abbott et~al.}{Abbott
  et~al.}{2020b}]{Abbott:2020niy}
Abbott R.,  et~al., 2020b, arXiv:2010.14527

\bibitem[\protect\citeauthoryear{Abbott et~al.}{Abbott
  et~al.}{2020a}]{Abbott:2020gyp}
Abbott R.,  et~al., 2020a, arXiv:2010.14533

\bibitem[\protect\citeauthoryear{Abbott et~al.}{Abbott
  et~al.}{2020c}]{Abbott:2020uma}
Abbott B.,  et~al., 2020c, \mn@doi [Astrophys. J. Lett.]
  {10.3847/2041-8213/ab75f5}, 892, L3

\bibitem[\protect\citeauthoryear{Abbott et~al.,}{Abbott
  et~al.}{2020d}]{Abbott_2020}
Abbott R.,  et~al., 2020d, \mn@doi [The Astrophysical Journal]
  {10.3847/2041-8213/ab960f}, 896, L44

\bibitem[\protect\citeauthoryear{Alford, Han  \& Prakash}{Alford
  et~al.}{2013}]{Alford:2013aca}
Alford M.~G.,  Han S.,   Prakash M.,  2013, \mn@doi [Phys. Rev. D]
  {10.1103/PhysRevD.88.083013}, 88, 083013

\bibitem[\protect\citeauthoryear{Antier et~al.}{Antier
  et~al.}{2020a}]{Antier:2019pzz}
Antier S.,  et~al., 2020a, \mn@doi [Mon. Not. Roy. Astron. Soc.]
  {10.1093/mnras/stz3142}, 492, 3904

\bibitem[\protect\citeauthoryear{{Antier} et~al.,}{{Antier}
  et~al.}{2020b}]{GRANDMA2020}
{Antier} S.,  et~al., 2020b, \mn@doi [\mnras] {10.1093/mnras/staa1846}, \href
  {https://ui.adsabs.harvard.edu/abs/2020MNRAS.497.5518A} {497, 5518}

\bibitem[\protect\citeauthoryear{Aubin et~al.,}{Aubin
  et~al.}{2020}]{Aubin:2020}
Aubin F.,  et~al., 2020, arXiv:2012.11512

\bibitem[\protect\citeauthoryear{{B. P. Abbott et al.}}{{B. P. Abbott et
  al.}}{2019}]{AbEA2019}
{B. P. Abbott et al.} 2019, \mn@doi [The Astrophysical Journal]
  {10.3847/1538-4357/ab0e8f}, 875, 161

\bibitem[\protect\citeauthoryear{Bauswein, Baumgarte  \& Janka}{Bauswein
  et~al.}{2013a}]{BaBa2013}
Bauswein A.,  Baumgarte T.~W.,   Janka H.-T.,  2013a, \mn@doi [Phys. Rev.
  Lett.] {10.1103/PhysRevLett.111.131101}, 111, 131101

\bibitem[\protect\citeauthoryear{Bauswein, Baumgarte  \& Janka}{Bauswein
  et~al.}{2013b}]{Bauswein:2013jpa}
Bauswein A.,  Baumgarte T.,   Janka H.~T.,  2013b, \mn@doi [Phys.Rev.Lett.]
  {10.1103/PhysRevLett.111.131101}, 111, 131101

\bibitem[\protect\citeauthoryear{{Bauswein et al.}}{{Bauswein et
  al.}}{2017}]{BaJu2017}
{Bauswein et al.} 2017, The Astrophysical Journal Letters, 850, L34

\bibitem[\protect\citeauthoryear{Bauswein, Blacker, Lioutas, Soultanis, Vijayan
   \& Stergioulas}{Bauswein et~al.}{2020}]{Bauswein:2020xlt}
Bauswein A.,  Blacker S.,  Lioutas G.,  Soultanis T.,  Vijayan V.,
  Stergioulas N.,  2020, arXiv: 2010.04461

\bibitem[\protect\citeauthoryear{Bellm et~al.,}{Bellm et~al.}{2018}]{Bellm2018}
Bellm E.~C.,  et~al., 2018, \mn@doi [Publications of the Astronomical Society
  of the Pacific] {10.1088/1538-3873/aaecbe}, 131, 018002

\bibitem[\protect\citeauthoryear{Berry, Mandel, Middleton  et~al.}{Berry
  et~al.}{2015}]{BeMa2015}
Berry C. P.~L.,  Mandel I.,  Middleton H.,   et~al., 2015, \mn@doi [Astrophys.
  J.] {10.1088/0004-637X/804/2/114}, 804, 114

\bibitem[\protect\citeauthoryear{Biscoveanu, Vitale  \& Haster}{Biscoveanu
  et~al.}{2019}]{Biscoveanu_2019}
Biscoveanu S.,  Vitale S.,   Haster C.-J.,  2019, \mn@doi [The Astrophysical
  Journal] {10.3847/2041-8213/ab479e}, 884, L32

\bibitem[\protect\citeauthoryear{{Bloemen}, {Groot}, {Nelemans}  \&
  {Klein-Wolt}}{{Bloemen} et~al.}{2015}]{BlGr2015}
{Bloemen} S.,  {Groot} P.,  {Nelemans} G.,   {Klein-Wolt} M.,  2015, in
  {Rucinski} S.~M.,  {Torres} G.,   {Zejda} M.,  eds,  Astronomical Society of
  the Pacific Conference Series Vol. 496, Living Together: Planets, Host Stars
  and Binaries. p.~254

\bibitem[\protect\citeauthoryear{Bogdanov et~al.}{Bogdanov
  et~al.}{2019}]{Bogdanov:2019ixe}
Bogdanov S.,  et~al., 2019, \mn@doi [Astrophys. J. Lett.]
  {10.3847/2041-8213/ab53eb}, 887, L25

\bibitem[\protect\citeauthoryear{Bovard, Martin, Guercilena, Arcones, Rezzolla
  \& Korobkin}{Bovard et~al.}{2017}]{Bovard:2017mvn}
Bovard L.,  Martin D.,  Guercilena F.,  Arcones A.,  Rezzolla L.,   Korobkin
  O.,  2017, \mn@doi [Phys. Rev.] {10.1103/PhysRevD.96.124005}, D96, 124005

\bibitem[\protect\citeauthoryear{Breschi, Perego, Bernuzzi, Del~Pozzo, Nedora,
  Radice  \& Vescovi}{Breschi et~al.}{2021}]{Breschi:2021tbm}
Breschi M.,  Perego A.,  Bernuzzi S.,  Del~Pozzo W.,  Nedora V.,  Radice D.,
  Vescovi D.,  2021, preprint arXiv:2101.01201

\bibitem[\protect\citeauthoryear{Breu \& Rezzolla}{Breu \&
  Rezzolla}{2016}]{10.1093/mnras/stw575}
Breu C.,  Rezzolla L.,  2016, \mn@doi [Monthly Notices of the Royal
  Astronomical Society] {10.1093/mnras/stw575}, 459, 646

\bibitem[\protect\citeauthoryear{Bulla}{Bulla}{2019}]{Bulla:2019muo}
Bulla M.,  2019, \mn@doi [Mon. Not. Roy. Astron. Soc.] {10.1093/mnras/stz2495},
  489, 5037

\bibitem[\protect\citeauthoryear{{Cannon} et~al.,}{{Cannon}
  et~al.}{2020}]{2020arXiv201005082C}
{Cannon} K.,  et~al., 2020, arXiv e-prints, \href
  {https://ui.adsabs.harvard.edu/abs/2020arXiv201005082C} {p. arXiv:2010.05082}

\bibitem[\protect\citeauthoryear{Capano et~al.,}{Capano
  et~al.}{2020}]{Capano:2019eae}
Capano C.~D.,  et~al., 2020, \mn@doi [Nature Astron.]
  {10.1038/s41550-020-1014-6}, 4, 625

\bibitem[\protect\citeauthoryear{Chatterjee, Ghosh, Brady, Kapadia, Miller,
  Nissanke  \& Pannarale}{Chatterjee et~al.}{2020}]{Chatterjee:2019avs}
Chatterjee D.,  Ghosh S.,  Brady P.~R.,  Kapadia S.~J.,  Miller A.~L.,
  Nissanke S.,   Pannarale F.,  2020, \mn@doi [Astrophys. J.]
  {10.3847/1538-4357/ab8dbe}, 896, 54

\bibitem[\protect\citeauthoryear{Chatziioannou \& Han}{Chatziioannou \&
  Han}{2020}]{Chatziioannou:2019yko}
Chatziioannou K.,  Han S.,  2020, \mn@doi [Phys. Rev. D]
  {10.1103/PhysRevD.101.044019}, 101, 044019

\bibitem[\protect\citeauthoryear{Chen, Essick, Vitale, Holz  \&
  Katsavounidis}{Chen et~al.}{2017}]{Chen2017}
Chen H.-Y.,  Essick R.,  Vitale S.,  Holz D.~E.,   Katsavounidis E.,  2017,
  \mn@doi [The Astrophysical Journal] {10.3847/1538-4357/835/1/31}, 835, 31

\bibitem[\protect\citeauthoryear{Christie, Lalakos, Tchekhovskoy, Fern\'andez,
  Foucart, Quataert  \& Kasen}{Christie et~al.}{2019}]{Christie:2019lim}
Christie I.~M.,  Lalakos A.,  Tchekhovskoy A.,  Fern\'andez R.,  Foucart F.,
  Quataert E.,   Kasen D.,  2019, \mn@doi [Mon. Not. Roy. Astron. Soc.]
  {10.1093/mnras/stz2552}, 490, 4811

\bibitem[\protect\citeauthoryear{Cornish \& Littenberg}{Cornish \&
  Littenberg}{2015}]{CoLi2015}
Cornish N.~J.,  Littenberg T.~B.,  2015, Classical and Quantum Gravity, 32,
  135012

\bibitem[\protect\citeauthoryear{Cornish, Littenberg, B\'ecsy, Chatziioannou,
  Clark, Ghonge  \& Millhouse}{Cornish et~al.}{2021}]{PhysRevD.103.044006}
Cornish N.~J.,  Littenberg T.~B.,  B\'ecsy B.,  Chatziioannou K.,  Clark J.~A.,
   Ghonge S.,   Millhouse M.,  2021, \mn@doi [Phys. Rev. D]
  {10.1103/PhysRevD.103.044006}, 103, 044006

\bibitem[\protect\citeauthoryear{Coughlin \& Dietrich}{Coughlin \&
  Dietrich}{2019}]{Coughlin:2019kqf}
Coughlin M.~W.,  Dietrich T.,  2019, \mn@doi [Phys. Rev. D]
  {10.1103/PhysRevD.100.043011}, 100, 043011

\bibitem[\protect\citeauthoryear{Coughlin, Dietrich, Kawaguchi, Smartt, Stubbs
  \& Ujevic}{Coughlin et~al.}{2017}]{CoDi2017}
Coughlin M.,  Dietrich T.,  Kawaguchi K.,  Smartt S.,  Stubbs C.,   Ujevic M.,
  2017, \mn@doi [Astrophys. J.] {10.3847/1538-4357/aa9114}, 849, 12

\bibitem[\protect\citeauthoryear{{Coughlin} et~al.,}{{Coughlin}
  et~al.}{2018a}]{CoDi2018}
{Coughlin} M.~W.,  et~al., 2018a, preprint, \href
  {http://adsabs.harvard.edu/abs/2018arXiv180509371C} {} (\mn@eprint {arXiv}
  {1805.09371})

\bibitem[\protect\citeauthoryear{Coughlin et~al.,}{Coughlin
  et~al.}{2018b}]{Coughlin:2018miv}
Coughlin M.~W.,  et~al., 2018b, \mn@doi [Monthly Notices of the Royal
  Astronomical Society] {10.1093/mnras/sty2174}, 480, 3871

\bibitem[\protect\citeauthoryear{Coughlin, Dietrich, Margalit  \&
  Metzger}{Coughlin et~al.}{2019a}]{Coughlin:2018fis}
Coughlin M.~W.,  Dietrich T.,  Margalit B.,   Metzger B.~D.,  2019a, \mn@doi
  [Mon. Not. Roy. Astron. Soc.] {10.1093/mnrasl/slz133}, 489, L91

\bibitem[\protect\citeauthoryear{Coughlin et~al.}{Coughlin
  et~al.}{2019b}]{Coughlin:2019xfb}
Coughlin M.~W.,  et~al., 2019b, \mn@doi [Astrophys. J.]
  {10.3847/2041-8213/ab4ad8}, 885, L19

\bibitem[\protect\citeauthoryear{{Coughlin} et~al.,}{{Coughlin}
  et~al.}{2020a}]{2020NatCo..11.4129C}
{Coughlin} M.~W.,  et~al., 2020a, \mn@doi [Nature Communications]
  {10.1038/s41467-020-17998-5}, \href
  {https://ui.adsabs.harvard.edu/abs/2020NatCo..11.4129C} {11, 4129}

\bibitem[\protect\citeauthoryear{Coughlin et~al.,}{Coughlin
  et~al.}{2020b}]{Coughlin:2019zqi}
Coughlin M.~W.,  et~al., 2020b, \mn@doi [Mon. Not. Roy. Astron. Soc.]
  {10.1093/mnras/stz3457}, 492, 863

\bibitem[\protect\citeauthoryear{Coughlin et~al.}{Coughlin
  et~al.}{2020c}]{Coughlin:2020fwx}
Coughlin M.~W.,  et~al., 2020c, \mn@doi [Mon. Not. Roy. Astron. Soc.]
  {10.1093/mnras/staa1925}, 497, 1181

\bibitem[\protect\citeauthoryear{{Coulter et al.}}{{Coulter et
  al.}}{2017}]{CoFo2017}
{Coulter et al.} 2017, \mn@doi [Science] {10.1126/science.aap9811}, 358, 1556

\bibitem[\protect\citeauthoryear{Dal~Canton et~al.}{Dal~Canton
  et~al.}{2014}]{Canton:2014ena}
Dal~Canton T.,  et~al., 2014, \mn@doi [Phys. Rev.]
  {10.1103/PhysRevD.90.082004}, D90, 082004

\bibitem[\protect\citeauthoryear{{Dal Canton}, {Nitz}, {Gadre}, {Davies},
  {Villa-Ortega}, {Dent}, {Harry}  \& {Xiao}}{{Dal Canton}
  et~al.}{2020}]{2020arXiv200807494D}
{Dal Canton} T.,  {Nitz} A.~H.,  {Gadre} B.,  {Davies} G.~S.,  {Villa-Ortega}
  V.,  {Dent} T.,  {Harry} I.,   {Xiao} L.,  2020, arXiv e-prints, \href
  {https://ui.adsabs.harvard.edu/abs/2020arXiv200807494D} {p. arXiv:2008.07494}

\bibitem[\protect\citeauthoryear{Dekany et~al.,}{Dekany
  et~al.}{2020}]{DeSm2018}
Dekany R.,  et~al., 2020, \mn@doi [Publications of the Astronomical Society of
  the Pacific] {10.1088/1538-3873/ab4ca2}, 132, 038001

\bibitem[\protect\citeauthoryear{Dietrich \& Ujevic}{Dietrich \&
  Ujevic}{2017}]{DiUj2017}
Dietrich T.,  Ujevic M.,  2017, \mn@doi [Class. Quant. Grav.]
  {10.1088/1361-6382/aa6bb0}, 34, 105014

\bibitem[\protect\citeauthoryear{Dietrich et~al.,}{Dietrich
  et~al.}{2018}]{Dietrich:2018phi}
Dietrich T.,  et~al., 2018, arXiv: 1806.01625

\bibitem[\protect\citeauthoryear{Dietrich, Coughlin, Pang, Bulla, Heinzel,
  Issa, Tews  \& Antier}{Dietrich et~al.}{2020}]{Dietrich:2020lps}
Dietrich T.,  Coughlin M.~W.,  Pang P. T.~H.,  Bulla M.,  Heinzel J.,  Issa L.,
   Tews I.,   Antier S.,  2020, \mn@doi [Science] {10.1126/science.abb4317},
  370, 1450

\bibitem[\protect\citeauthoryear{Douchin \& Haensel}{Douchin \&
  Haensel}{2001}]{Douchin:2001sv}
Douchin F.,  Haensel P.,  2001, Astron. Astrophys., 380, 151

\bibitem[\protect\citeauthoryear{Essick \& Landry}{Essick \&
  Landry}{2020}]{Essick:2020ghc}
Essick R.,  Landry P.,  2020, \mn@doi [Astrophys. J.]
  {10.3847/1538-4357/abbd3b}, 904, 80

\bibitem[\protect\citeauthoryear{Essick, Vitale, Katsavounidis, Vedovato  \&
  Klimenko}{Essick et~al.}{2015}]{EsVi2015}
Essick R.,  Vitale S.,  Katsavounidis E.,  Vedovato G.,   Klimenko S.,  2015,
  The Astrophysical Journal, 800, 81

\bibitem[\protect\citeauthoryear{Essick, Landry  \& Holz}{Essick
  et~al.}{2020a}]{Essick:2019ldf}
Essick R.,  Landry P.,   Holz D.~E.,  2020a, \mn@doi [Phys. Rev. D]
  {10.1103/PhysRevD.101.063007}, 101, 063007

\bibitem[\protect\citeauthoryear{Essick, Tews, Landry, Reddy  \& Holz}{Essick
  et~al.}{2020b}]{PhysRevC.102.055803}
Essick R.,  Tews I.,  Landry P.,  Reddy S.,   Holz D.~E.,  2020b, \mn@doi
  [Phys. Rev. C] {10.1103/PhysRevC.102.055803}, 102, 055803

\bibitem[\protect\citeauthoryear{{Fairhurst}}{{Fairhurst}}{2009}]{Fair2009}
{Fairhurst} S.,  2009, \mn@doi [New J. Phys.] {10.1088/1367-2630/11/12/123006},
  \href {http://adsabs.harvard.edu/abs/2009NJPh...11l3006F} {11, 123006}

\bibitem[\protect\citeauthoryear{Fairhurst}{Fairhurst}{2011}]{Fair2011}
Fairhurst S.,  2011, \mn@doi [Class. Quant. Grav.]
  {10.1088/0264-9381/28/10/105021}, 28, 105021

\bibitem[\protect\citeauthoryear{Fern\'andez, Kasen, Metzger  \&
  Quataert}{Fern\'andez et~al.}{2015}]{Fernandez:2014cna}
Fern\'andez R.,  Kasen D.,  Metzger B.~D.,   Quataert E.,  2015, \mn@doi [Mon.
  Not. Roy. Astron. Soc.] {10.1093/mnras/stu2112}, 446, 750

\bibitem[\protect\citeauthoryear{Fern\'andez, Tchekhovskoy, Quataert, Foucart
  \& Kasen}{Fern\'andez et~al.}{2019}]{Fernandez:2018kax}
Fern\'andez R.,  Tchekhovskoy A.,  Quataert E.,  Foucart F.,   Kasen D.,  2019,
  \mn@doi [Mon. Not. Roy. Astron. Soc.] {10.1093/mnras/sty2932}, 482, 3373

\bibitem[\protect\citeauthoryear{Fishbach, Farr  \& Holz}{Fishbach
  et~al.}{2020a}]{Fishbach_2020}
Fishbach M.,  Farr W.~M.,   Holz D.~E.,  2020a, \mn@doi [The Astrophysical
  Journal] {10.3847/2041-8213/ab77c9}, 891, L31

\bibitem[\protect\citeauthoryear{Fishbach, Essick  \& Holz}{Fishbach
  et~al.}{2020b}]{Fishbach:2020ryj}
Fishbach M.,  Essick R.,   Holz D.~E.,  2020b, \mn@doi [Astrophys. J. Lett.]
  {10.3847/2041-8213/aba7b6}, 899, L8

\bibitem[\protect\citeauthoryear{Foucart}{Foucart}{2012}]{Foucart:2012nc}
Foucart F.,  2012, \mn@doi [Phys. Rev. D] {10.1103/PhysRevD.86.124007}, 86,
  124007

\bibitem[\protect\citeauthoryear{Foucart}{Foucart}{2020}]{Foucart:2020ats}
Foucart F.,  2020, \mn@doi [Front. Astron. Space Sci.]
  {10.3389/fspas.2020.00046}, 7, 46

\bibitem[\protect\citeauthoryear{Foucart, Hinderer  \& Nissanke}{Foucart
  et~al.}{2018}]{Foucart:2018rjc}
Foucart F.,  Hinderer T.,   Nissanke S.,  2018, \mn@doi [Phys. Rev. D]
  {10.1103/PhysRevD.98.081501}, 98, 081501

\bibitem[\protect\citeauthoryear{Foucart, Duez, Kidder, Nissanke, Pfeiffer  \&
  Scheel}{Foucart et~al.}{2019}]{Foucart:2019bxj}
Foucart F.,  Duez M.,  Kidder L.,  Nissanke S.,  Pfeiffer H.,   Scheel M.,
  2019, \mn@doi [Phys. Rev. D] {10.1103/PhysRevD.99.103025}, 99, 103025

\bibitem[\protect\citeauthoryear{Goldstein et~al.}{Goldstein
  et~al.}{2017}]{Goldstein:2017mmi}
Goldstein A.,  et~al., 2017, \mn@doi [Astrophys. J.]
  {10.3847/2041-8213/aa8f41}, 848, L14

\bibitem[\protect\citeauthoryear{Gompertz et~al.,}{Gompertz
  et~al.}{2020}]{GoCu2020}
Gompertz B.~P.,  et~al., 2020, \mn@doi [Monthly Notices of the Royal
  Astronomical Society] {10.1093/mnras/staa1845}, 497, 726

\bibitem[\protect\citeauthoryear{Goriely, Bauswein  \& Janka}{Goriely
  et~al.}{2011}]{Goriely:2011vg}
Goriely S.,  Bauswein A.,   Janka H.-T.,  2011, \mn@doi [Astrophys.J.]
  {10.1088/2041-8205/738/2/L32}, 738, L32

\bibitem[\protect\citeauthoryear{Graham et~al.,}{Graham
  et~al.}{2019}]{Graham2018}
Graham M.~J.,  et~al., 2019, \mn@doi [Publications of the Astronomical Society
  of the Pacific] {10.1088/1538-3873/ab006c}, 131, 078001

\bibitem[\protect\citeauthoryear{Grossman, Korobkin, Rosswog  \&
  Piran}{Grossman et~al.}{2014}]{Grossman:2013lqa}
Grossman D.,  Korobkin O.,  Rosswog S.,   Piran T.,  2014, \mn@doi [Mon. Not.
  Roy. Astron. Soc.] {10.1093/mnras/stt2503}, 439, 757

\bibitem[\protect\citeauthoryear{Grover, Fairhurst, Farr  et~al.}{Grover
  et~al.}{2014}]{Grover:2013}
Grover K.,  Fairhurst S.,  Farr B.~F.,   et~al., 2014, \mn@doi [Phys. Rev.]
  {10.1103/PhysRevD.89.042004}, D89, 042004

\bibitem[\protect\citeauthoryear{Han, Mamun, Lalit, Constantinou  \&
  Prakash}{Han et~al.}{2019}]{Han:2019bub}
Han S.,  Mamun M.,  Lalit S.,  Constantinou C.,   Prakash M.,  2019, \mn@doi
  [Phys. Rev. D] {10.1103/PhysRevD.100.103022}, 100, 103022

\bibitem[\protect\citeauthoryear{Heinzel et~al.,}{Heinzel
  et~al.}{2020}]{Heinzel:2020qlt}
Heinzel J.,  et~al., 2020, arXiv:2010.10746

\bibitem[\protect\citeauthoryear{Hinderer et~al.}{Hinderer
  et~al.}{2019}]{Hinderer:2018pei}
Hinderer T.,  et~al., 2019, \mn@doi [Phys. Rev. D]
  {10.1103/PhysRevD.100.063021}, 100, 06321

\bibitem[\protect\citeauthoryear{Hooper, Chung, Luan, Blair, Chen  \&
  Wen}{Hooper et~al.}{2012}]{Hooper:2011rb}
Hooper S.,  Chung S.~K.,  Luan J.,  Blair D.,  Chen Y.,   Wen L.,  2012,
  \mn@doi [Phys. Rev. D] {10.1103/PhysRevD.86.024012}, 86, 024012

\bibitem[\protect\citeauthoryear{Ivezic, Tyson, Allsman, Andrew  \&
  Angel}{Ivezic et~al.}{2019}]{Ivezic2014}
Ivezic Z.,  Tyson J.~A.,  Allsman R.,  Andrew J.,   Angel R.,  2019, \mn@doi
  [\apj] {10.3847/1538-4357/ab042c}, \href
  {https://ui.adsabs.harvard.edu/abs/2019ApJ...873..111I} {873, 111}

\bibitem[\protect\citeauthoryear{Kapadia et~al.}{Kapadia
  et~al.}{2020}]{Kapadia:2019uut}
Kapadia S.~J.,  et~al., 2020, \mn@doi [Class. Quant. Grav.]
  {10.1088/1361-6382/ab5f2d}, 37, 045007

\bibitem[\protect\citeauthoryear{Kasen, Metzger, Barnes, Quataert  \&
  Ramirez-Ruiz}{Kasen et~al.}{2017}]{KaMe2017}
Kasen D.,  Metzger B.,  Barnes J.,  Quataert E.,   Ramirez-Ruiz E.,  2017,
  Nature, 551, 80 EP

\bibitem[\protect\citeauthoryear{Kawaguchi, Kyutoku, Shibata  \&
  Tanaka}{Kawaguchi et~al.}{2016}]{Kawaguchi:2016ana}
Kawaguchi K.,  Kyutoku K.,  Shibata M.,   Tanaka M.,  2016, \mn@doi [Astrophys.
  J.] {10.3847/0004-637X/825/1/52}, 825, 52

\bibitem[\protect\citeauthoryear{Kawaguchi, Shibata  \& Tanaka}{Kawaguchi
  et~al.}{2019}]{Kawaguchi:2019nju}
Kawaguchi K.,  Shibata M.,   Tanaka M.,  2019, \mn@doi [arXiv:1908.05815]
  {10.3847/1538-4357/ab61f6}

\bibitem[\protect\citeauthoryear{Klimenko, Yakushin, Mercer  \&
  Mitselmakher}{Klimenko et~al.}{2008}]{Klimenko:2008fu}
Klimenko S.,  Yakushin I.,  Mercer A.,   Mitselmakher G.,  2008, \mn@doi
  [Class. Quant. Grav.] {10.1088/0264-9381/25/11/114029}, 25, 114029

\bibitem[\protect\citeauthoryear{Klimenko et~al.,}{Klimenko
  et~al.}{2016}]{KlVe2016}
Klimenko S.,  et~al., 2016, \mn@doi [Phys. Rev. D]
  {10.1103/PhysRevD.93.042004}, 93, 042004

\bibitem[\protect\citeauthoryear{Kr\"uger \& Foucart}{Kr\"uger \&
  Foucart}{2020}]{KrFo2020}
Kr\"uger C.~J.,  Foucart F.,  2020, \mn@doi [Phys. Rev. D]
  {10.1103/PhysRevD.101.103002}, 101, 103002

\bibitem[\protect\citeauthoryear{Krüger \& Foucart}{Krüger \&
  Foucart}{2020}]{Kruger:2020gig}
Krüger C.~J.,  Foucart F.,  2020, \mn@doi [Phys. Rev. D]
  {10.1103/PhysRevD.101.103002}, 101, 103002

\bibitem[\protect\citeauthoryear{{LIGO Scientific Collaboration} \& {Virgo
  Collaboration}}{{LIGO Scientific Collaboration} \& {Virgo
  Collaboration}}{2019a}]{2019GCN.24168....1L}
{LIGO Scientific Collaboration} {Virgo Collaboration} 2019a, GRB Coordinates
  Network, \href {https://ui.adsabs.harvard.edu/abs/2019GCN.24168....1L}
  {24168, 1}

\bibitem[\protect\citeauthoryear{{LIGO Scientific Collaboration} \& {Virgo
  Collaboration}}{{LIGO Scientific Collaboration} \& {Virgo
  Collaboration}}{2019b}]{2019GCN.25324....1L}
{LIGO Scientific Collaboration} {Virgo Collaboration} 2019b, GRB Coordinates
  Network, \href {https://ui.adsabs.harvard.edu/abs/2019GCN.25324....1L}
  {25324, 1}

\bibitem[\protect\citeauthoryear{Landry \& Essick}{Landry \&
  Essick}{2019}]{PhysRevD.99.084049}
Landry P.,  Essick R.,  2019, \mn@doi [Phys. Rev. D]
  {10.1103/PhysRevD.99.084049}, 99, 084049

\bibitem[\protect\citeauthoryear{Landry, Essick  \& Chatziioannou}{Landry
  et~al.}{2020}]{PhysRevD.101.123007}
Landry P.,  Essick R.,   Chatziioannou K.,  2020, \mn@doi [Phys. Rev. D]
  {10.1103/PhysRevD.101.123007}, 101, 123007

\bibitem[\protect\citeauthoryear{Lange, O'Shaughnessy  \& Rizzo}{Lange
  et~al.}{2018}]{Lange:2018pyp}
Lange J.,  O'Shaughnessy R.,   Rizzo M.,  2018, arXiv:1805.10457

\bibitem[\protect\citeauthoryear{Lattimer \& Prakash}{Lattimer \&
  Prakash}{2001}]{Lattimer_2001}
Lattimer J.~M.,  Prakash M.,  2001, \mn@doi [The Astrophysical Journal]
  {10.1086/319702}, 550, 426

\bibitem[\protect\citeauthoryear{{Lattimer} \& {Schramm}}{{Lattimer} \&
  {Schramm}}{1974}]{LaSc1974}
{Lattimer} J.~M.,  {Schramm} D.~N.,  1974, \mn@doi [\apjl] {10.1086/181612},
  \href {http://adsabs.harvard.edu/abs/1974ApJ...192L.145L} {192, L145}

\bibitem[\protect\citeauthoryear{Li \& Paczynski}{Li \&
  Paczynski}{1998}]{LiPa1998}
Li L.-X.,  Paczynski B.,  1998, The Astrophysical Journal Letters, 507, L59

\bibitem[\protect\citeauthoryear{Lindblom}{Lindblom}{1998}]{PhysRevD.58.024008}
Lindblom L.,  1998, \mn@doi [Phys. Rev. D] {10.1103/PhysRevD.58.024008}, 58,
  024008

\bibitem[\protect\citeauthoryear{Lindblom}{Lindblom}{2010}]{Lindblom:2010bb}
Lindblom L.,  2010, \mn@doi [Phys. Rev. D] {10.1103/PhysRevD.82.103011}, 82,
  103011

\bibitem[\protect\citeauthoryear{{Lindblom} \& {Indik}}{{Lindblom} \&
  {Indik}}{2012}]{2012PhRvD..86h4003L}
{Lindblom} L.,  {Indik} N.~M.,  2012, \mn@doi [\prd]
  {10.1103/PhysRevD.86.084003}, \href
  {https://ui.adsabs.harvard.edu/abs/2012PhRvD..86h4003L} {86, 084003}

\bibitem[\protect\citeauthoryear{Lindblom \& Indik}{Lindblom \&
  Indik}{2014}]{Lindblom:2013kra}
Lindblom L.,  Indik N.~M.,  2014, \mn@doi [Phys. Rev. D]
  {10.1103/PhysRevD.89.064003}, 89, 064003

\bibitem[\protect\citeauthoryear{Lynch, Vitale, Essick, Katsavounidis  \&
  Robinet}{Lynch et~al.}{2017}]{PhysRevD.95.104046}
Lynch R.,  Vitale S.,  Essick R.,  Katsavounidis E.,   Robinet F.,  2017,
  \mn@doi [Phys. Rev. D] {10.1103/PhysRevD.95.104046}, 95, 104046

\bibitem[\protect\citeauthoryear{Mandel}{Mandel}{2010}]{PhysRevD.81.084029}
Mandel I.,  2010, \mn@doi [Phys. Rev. D] {10.1103/PhysRevD.81.084029}, 81,
  084029

\bibitem[\protect\citeauthoryear{Margalit \& Metzger}{Margalit \&
  Metzger}{2019}]{Margalit:2019dpi}
Margalit B.,  Metzger B.~D.,  2019, \mn@doi [Astrophys. J. Lett.]
  {10.3847/2041-8213/ab2ae2}, 880, L15

\bibitem[\protect\citeauthoryear{Masci et~al.,}{Masci et~al.}{2018}]{MaLa2018}
Masci F.~J.,  et~al., 2018, \mn@doi [Publications of the Astronomical Society
  of the Pacific] {10.1088/1538-3873/aae8ac}, 131, 018003

\bibitem[\protect\citeauthoryear{{Metzger} et~al.,}{{Metzger}
  et~al.}{2010}]{MeMa2010}
{Metzger} B.~D.,  et~al., 2010, \mn@doi [Monthly Notices of the Royal
  Astronomical Society] {10.1111/j.1365-2966.2010.16864.x}, \href
  {http://adsabs.harvard.edu/abs/2010MNRAS.406.2650M} {406, 2650}

\bibitem[\protect\citeauthoryear{Miller et~al.}{Miller
  et~al.}{2019}]{Miller:2019cac}
Miller M.,  et~al., 2019, \mn@doi [Astrophys. J. Lett.]
  {10.3847/2041-8213/ab50c5}, 887, L24

\bibitem[\protect\citeauthoryear{Morgan, Kaiser, Moreau, Anderson  \&
  Burgett}{Morgan et~al.}{2012}]{MoKa2012}
Morgan J.~S.,  Kaiser N.,  Moreau V.,  Anderson D.,   Burgett W.,  2012,
  \mn@doi [Proc. SPIE Int. Soc. Opt. Eng.] {10.1117/12.926646}, 8444, 0H

\bibitem[\protect\citeauthoryear{Most, Papenfort, Weih  \& Rezzolla}{Most
  et~al.}{2020}]{Most:2020bba}
Most E.~R.,  Papenfort L.~J.,  Weih L.~R.,   Rezzolla L.,  2020, \mn@doi [Mon.
  Not. Roy. Astron. Soc.] {10.1093/mnrasl/slaa168}, 499, L82

\bibitem[\protect\citeauthoryear{Nedora et~al.,}{Nedora
  et~al.}{2020}]{Nedora:2020qtd}
Nedora V.,  et~al., 2020, {preprint arXiv:2011.11110}

\bibitem[\protect\citeauthoryear{{Nicholl}, {Margalit}, {Schmidt}, {Smith},
  {Ridley}  \& {Nuttall}}{{Nicholl} et~al.}{2021}]{2021arXiv210202229N}
{Nicholl} M.,  {Margalit} B.,  {Schmidt} P.,  {Smith} G.~P.,  {Ridley} E.~J.,
  {Nuttall} J.,  2021, arXiv e-prints, \href
  {https://ui.adsabs.harvard.edu/abs/2021arXiv210202229N} {p. arXiv:2102.02229}

\bibitem[\protect\citeauthoryear{Pang, Dietrich, Tews  \& Van Den~Broeck}{Pang
  et~al.}{2020}]{Pang:2020ilf}
Pang P.~T.,  Dietrich T.,  Tews I.,   Van Den~Broeck C.,  2020, \mn@doi [Phys.
  Rev. Res.] {10.1103/PhysRevResearch.2.033514}, 2, 033514

\bibitem[\protect\citeauthoryear{Pankow, Brady, Ochsner  \&
  O'Shaughnessy}{Pankow et~al.}{2015}]{Pankow:2015cra}
Pankow C.,  Brady P.,  Ochsner E.,   O'Shaughnessy R.,  2015, \mn@doi [Phys.
  Rev.] {10.1103/PhysRevD.92.023002}, D92, 023002

\bibitem[\protect\citeauthoryear{Pannarale \& Ohme}{Pannarale \&
  Ohme}{2014}]{Pannarale:2014rea}
Pannarale F.,  Ohme F.,  2014, \mn@doi [Astrophys. J.]
  {10.1088/2041-8205/791/1/L7}, 791, L7

\bibitem[\protect\citeauthoryear{Perego, Radice  \& Bernuzzi}{Perego
  et~al.}{2017}]{Perego:2017wtu}
Perego A.,  Radice D.,   Bernuzzi S.,  2017, \mn@doi [Astrophys. J.]
  {10.3847/2041-8213/aa9ab9}, 850, L37

\bibitem[\protect\citeauthoryear{Piran, Nakar  \& Rosswog}{Piran
  et~al.}{2013}]{PiNa2013}
Piran T.,  Nakar E.,   Rosswog S.,  2013, \mn@doi [Monthly Notices of the Royal
  Astronomical Society] {10.1093/mnras/stt037}, 430, 2121

\bibitem[\protect\citeauthoryear{Raaijmakers et~al.}{Raaijmakers
  et~al.}{2020}]{Raaijmakers:2019dks}
Raaijmakers G.,  et~al., 2020, \mn@doi [Astrophys. J. Lett.]
  {10.3847/2041-8213/ab822f}, 893, L21

\bibitem[\protect\citeauthoryear{Raaijmakers et~al.}{Raaijmakers
  et~al.}{2021}]{Raaijmakers:2021slr}
Raaijmakers G.,  et~al., 2021, preprint arXiv:2102.11569

\bibitem[\protect\citeauthoryear{Radice \& Dai}{Radice \&
  Dai}{2018}]{Radice:2018ozg}
Radice D.,  Dai L.,  2018, arXiv:1810.12917

\bibitem[\protect\citeauthoryear{Radice, Perego, Hotokezaka, Fromm, Bernuzzi
  \& Roberts}{Radice et~al.}{2018a}]{Radice:2018pdn}
Radice D.,  Perego A.,  Hotokezaka K.,  Fromm S.~A.,  Bernuzzi S.,   Roberts
  L.~F.,  2018a, arXiv: 1809.11161

\bibitem[\protect\citeauthoryear{{Radice et al.}}{{Radice et
  al.}}{2018b}]{RaPe2018}
{Radice et al.} 2018b, The Astrophysical Journal Letters, 852, L29

\bibitem[\protect\citeauthoryear{Riley et~al.}{Riley
  et~al.}{2019}]{Riley:2019yda}
Riley T.~E.,  et~al., 2019, \mn@doi [Astrophys. J. Lett.]
  {10.3847/2041-8213/ab481c}, 887, L21

\bibitem[\protect\citeauthoryear{Rosswog, Korobkin, Arcones, Thielemann  \&
  Piran}{Rosswog et~al.}{2014}]{Rosswog:2013kqa}
Rosswog S.,  Korobkin O.,  Arcones A.,  Thielemann F.,   Piran T.,  2014,
  \mn@doi [Mon. Not. Roy. Astron. Soc.] {10.1093/mnras/stt2502}, 439, 744

\bibitem[\protect\citeauthoryear{{R{\"o}ver}, {Meyer}, {Guidi}, {Vicer{\'e}}
  \& {Christensen}}{{R{\"o}ver} et~al.}{2007}]{Rover2007a}
{R{\"o}ver} C.,  {Meyer} R.,  {Guidi} G.~M.,  {Vicer{\'e}} A.,   {Christensen}
  N.,  2007, \mn@doi [Classical and Quantum Gravity]
  {10.1088/0264-9381/24/19/S23}, \href
  {http://cdsads.u-strasbg.fr/abs/2007CQGra..24S.607R} {24, S607}

\bibitem[\protect\citeauthoryear{Sachdev et~al.}{Sachdev
  et~al.}{2019}]{Sachdev:2019vvd}
Sachdev S.,  et~al., 2019, arXiv:1901.08580

\bibitem[\protect\citeauthoryear{Salafia, Colpi, Branchesi, Chassande-Mottin,
  Ghirlanda, Ghisellini  \& Vergani}{Salafia et~al.}{2017}]{SoCo2017}
Salafia O.~S.,  Colpi M.,  Branchesi M.,  Chassande-Mottin E.,  Ghirlanda G.,
  Ghisellini G.,   Vergani S.~D.,  2017, The Astrophysical Journal, 846, 62

\bibitem[\protect\citeauthoryear{Savchenko et~al.}{Savchenko
  et~al.}{2017}]{Savchenko:2017ffs}
Savchenko V.,  et~al., 2017, \mn@doi [Astrophys. J.]
  {10.3847/2041-8213/aa8f94}, 848, L15

\bibitem[\protect\citeauthoryear{Schnittman, Dal~Canton, Camp, Tsang  \&
  Kelly}{Schnittman et~al.}{2018}]{Schnittman:2017nhg}
Schnittman J.~D.,  Dal~Canton T.,  Camp J.,  Tsang D.,   Kelly B.~J.,  2018,
  \mn@doi [Astrophys. J.] {10.3847/1538-4357/aaa08b}, 853, 123

\bibitem[\protect\citeauthoryear{Shibata \& Taniguchi}{Shibata \&
  Taniguchi}{2011}]{Shibata:2011jka}
Shibata M.,  Taniguchi K.,  2011, \mn@doi [Living Rev. Rel.]
  {10.12942/lrr-2011-6}, 14, 6

\bibitem[\protect\citeauthoryear{Shibata, Fujibayashi, Hotokezaka, Kiuchi,
  Kyutoku, Sekiguchi  \& Tanaka}{Shibata et~al.}{2017}]{Shibata:2017xdx}
Shibata M.,  Fujibayashi S.,  Hotokezaka K.,  Kiuchi K.,  Kyutoku K.,
  Sekiguchi Y.,   Tanaka M.,  2017, \mn@doi [Phys. Rev.]
  {10.1103/PhysRevD.96.123012}, D96, 123012

\bibitem[\protect\citeauthoryear{Sidery, Aylott, Christensen  et~al.}{Sidery
  et~al.}{2014}]{SiAy2014}
Sidery T.,  Aylott B.,  Christensen N.,   et~al., 2014, \mn@doi [Phys. Rev.]
  {10.1103/PhysRevD.89.084060}, D89, 084060

\bibitem[\protect\citeauthoryear{Siegel \& Metzger}{Siegel \&
  Metzger}{2018}]{Siegel:2017jug}
Siegel D.~M.,  Metzger B.~D.,  2018, \mn@doi [Astrophys. J.]
  {10.3847/1538-4357/aabaec}, 858, 52

\bibitem[\protect\citeauthoryear{Singer \& Price}{Singer \&
  Price}{2016}]{SiPr2016}
Singer L.~P.,  Price L.~R.,  2016, \mn@doi [Phys. Rev. D]
  {10.1103/PhysRevD.93.024013}, 93, 024013

\bibitem[\protect\citeauthoryear{Singer, Price, Farr  et~al.}{Singer
  et~al.}{2014}]{SiPr2014}
Singer L.~P.,  Price L.~R.,  Farr B.,   et~al., 2014, \mn@doi [Astrophys. J.]
  {10.1088/0004-637X/795/2/105}, 795, 105

\bibitem[\protect\citeauthoryear{{Singer et al.}}{{Singer et
  al.}}{2016}]{Si2016}
{Singer et al.} 2016, The Astrophysical Journal Letters, 829, L15

\bibitem[\protect\citeauthoryear{{Smartt et al.}}{{Smartt et
  al.}}{2017}]{SmCh2017}
{Smartt et al.} 2017, Nature, 551, 75 EP

\bibitem[\protect\citeauthoryear{Smith, Ashton, Vajpeyi  \& Talbot}{Smith
  et~al.}{2020}]{Smith:2019ucc}
Smith R. J.~E.,  Ashton G.,  Vajpeyi A.,   Talbot C.,  2020, \mn@doi [Mon. Not.
  Roy. Astron. Soc.] {10.1093/mnras/staa2483}, 498, 4492

\bibitem[\protect\citeauthoryear{{Sridhar}, {Zrake}, {Metzger}, {Sironi}  \&
  {Giannios}}{{Sridhar} et~al.}{2021}]{2021MNRAS.501.3184S}
{Sridhar} N.,  {Zrake} J.,  {Metzger} B.~D.,  {Sironi} L.,   {Giannios} D.,
  2021, \mn@doi [\mnras] {10.1093/mnras/staa3794}, \href
  {https://ui.adsabs.harvard.edu/abs/2021MNRAS.501.3184S} {501, 3184}

\bibitem[\protect\citeauthoryear{Sutton et~al.,}{Sutton
  et~al.}{2010}]{Sutton_2010}
Sutton P.~J.,  et~al., 2010, \mn@doi [New Journal of Physics]
  {10.1088/1367-2630/12/5/053034}, 12, 053034

\bibitem[\protect\citeauthoryear{Tan, Noronha-Hostler  \& Yunes}{Tan
  et~al.}{2020}]{Tan:2020ics}
Tan H.,  Noronha-Hostler J.,   Yunes N.,  2020, \mn@doi [Phys. Rev. Lett.]
  {10.1103/PhysRevLett.125.261104}, 125, 261104

\bibitem[\protect\citeauthoryear{Tanaka \& Hotokezaka}{Tanaka \&
  Hotokezaka}{2013}]{Tanaka:2013ana}
Tanaka M.,  Hotokezaka K.,  2013, \mn@doi [Astrophys.J.]
  {10.1088/0004-637X/775/2/113}, 775, 113

\bibitem[\protect\citeauthoryear{Tews, Pang, Dietrich, Coughlin, Antier, Bulla,
  Heinzel  \& Issa}{Tews et~al.}{2020}]{Tews:2020ylw}
Tews I.,  Pang P.~T.,  Dietrich T.,  Coughlin M.~W.,  Antier S.,  Bulla M.,
  Heinzel J.,   Issa L.,  2020, arXiv:2007.06057

\bibitem[\protect\citeauthoryear{Tonry et~al.,}{Tonry et~al.}{2018}]{ToDe2018}
Tonry J.~L.,  et~al., 2018, Publications of the Astronomical Society of the
  Pacific, 130, 064505

\bibitem[\protect\citeauthoryear{Veitch et~al.}{Veitch
  et~al.}{2015}]{Veitch:2014wba}
Veitch J.,  et~al., 2015, \mn@doi [Phys. Rev.] {10.1103/PhysRevD.91.042003},
  D91, 042003

\bibitem[\protect\citeauthoryear{Wen \& Chen}{Wen \& Chen}{2010}]{WeCh2010}
Wen L.,  Chen Y.,  2010, \mn@doi [Phys. Rev.] {10.1103/PhysRevD.81.082001},
  D81, 082001

\makeatother
\end{thebibliography}


\bsp	
\label{lastpage}
\end{document}